\documentclass[aps,prd,
superscriptaddress,
showpacs,
preprintnumbers,
nofootinbib,
bibnotes,
amsmath,
amssymb,
twocolumn,
floatfix
]{revtex4-2}
\usepackage[utf8]{inputenc}
\usepackage{array,bbold}
\usepackage{todonotes}
\setuptodonotes{inline}
\usepackage[normalem]{ulem}
\newcolumntype{L}[1]{>{\raggedright\let\newline\\\arraybackslash\hspace{0pt}}m{#1}}
\newcolumntype{C}[1]{>{\centering\let\newline\\\arraybackslash\hspace{0pt}}m{#1}}
\newcolumntype{R}[1]{>{\raggedleft\let\newline\\\arraybackslash\hspace{0pt}}m{#1}}

\newcommand{\Hop}{H}
\DeclareMathOperator{\SU}{SU}
\DeclareMathOperator{\End}{End}
\DeclareMathOperator{\Hom}{Hom}

\usepackage[caption=false]{subfig}
\usepackage{listings}

\usepackage{xcolor}

\definecolor{codegreen}{rgb}{0,0.6,0}
\definecolor{codegray}{rgb}{0.5,0.5,0.5}
\definecolor{codepurple}{rgb}{0.58,0,0.82}
\definecolor{backcolour}{rgb}{0.95,0.95,0.92}

\lstdefinestyle{mystyle}{
    backgroundcolor=\color{backcolour},   
    commentstyle=\color{codegreen},
    keywordstyle=\color{magenta},
    numberstyle=\tiny\color{codegray},
    stringstyle=\color{codepurple},
    basicstyle=\ttfamily\footnotesize,
    breakatwhitespace=false,         
    breaklines=true,                 
    captionpos=b,                    
    keepspaces=true,                 
    numbers=left,                    
    numbersep=5pt,                  
    showspaces=false,                
    showstringspaces=false,
    showtabs=false,                  
    tabsize=2
}

\usepackage{graphicx}
\usepackage{dcolumn}
\usepackage{bm}
\usepackage{verbatim}
\usepackage{amsfonts}
\usepackage{longtable}
\usepackage{natbib}
\usepackage{hyperref}
\usepackage{dcolumn}

\newcommand\coarse[1]{\tilde{#1}}
\newcommand\1{\mathbb{1}}
\newcommand\C{\mathbb{C}}
\newcommand\N{\mathbb{N}}
\newcommand\R{\mathbb{R}}
\renewcommand\phi\varphi

\newcommand\regensburg{Department of Physics, University of Regensburg, 93040 Regensburg, Germany}

\begin{document}
\title{Gauge-equivariant neural networks as preconditioners in lattice QCD}

\author{C.~Lehner}\thanks{Corresponding author}\email{christoph.lehner@ur.de}\affiliation{\regensburg}
\author{T.~Wettig}\affiliation{\regensburg}

\date{\today}

\begin{abstract} 
  We demonstrate that a state-of-the art multi-grid preconditioner can be learned efficiently by gauge-equivariant neural networks.  We show that the models require minimal re-training on different gauge configurations of the same gauge ensemble and to a large extent remain efficient under modest modifications of ensemble parameters.  We also demonstrate that important paradigms such as communication avoidance are straightforward to implement in this framework.
\end{abstract}

\preprint{}

\keywords{machine learning, lattice QCD}

\maketitle

\section{Introduction}

Our current understanding of nature at the most fundamental level is to a large extent based on quantum field theories. In particle physics, Quantum Chromodynamics (QCD) explains, for example, how the proton is made up of smaller constituents, quarks and gluons. To describe current and future experiments, and to search for physics beyond the Standard Model, we need to be able to solve QCD to high precision. Lattice QCD constitutes a systematically improvable tool to solve QCD in the nonperturbative regime by numerically simulating the theory on a finite space-time lattice. It has evolved over more than four decades and is now of direct phenomenological relevance, see \cite{USQCD:2022mmc} and references therein. It is also very compute-intensive and employs the largest supercomputers worldwide \cite{Boyle:2022ncb}. Therefore much research is focused on improving the algorithms that dominate the run time of these simulations.

The most time-consuming element, both in the generation of gauge-field configurations and in the computation of physical observables, is typically the solution of the Dirac equation in the presence of a given gauge field. For physical values of the light quark masses and large lattice volumes, the condition number of the matrix representing the Dirac operator becomes very large, and consequently very sophisticated methods are required to solve the Dirac equation in a feasible time frame. The current state of the art is to use a suitable preconditioner inside a Krylov subspace solver. The construction of the preconditioner is a complicated problem whose solution requires deep knowledge of the underlying physics. The aim of this paper is to reformulate the problem in the language of gauge-equivariant neural networks and to show that such networks can learn the general paradigms of state-of-the-art preconditioners and efficiently reduce the iteration count of the outer solver. We also provide a flexible implementation interface in the Grid Python Toolkit (GPT) \cite{GPT} that allows for experimentation and further studies.

We briefly relate this paper to previous work. We will concentrate on multi-grid preconditioners \cite{Brannick:2007ue,Babich:2010qb,Frommer:2013fsa,Brannick:2014vda,Brower:2018ymy,Brower:2020xmc,Boyle:2021wcf} and refer to \cite{trottenberg2000multigrid} for an introduction. The idea of learning the elements of multi-grid preconditioners with neural networks has been pursued in a number of earlier publications, see, e.g., \cite{Katrutsa:2017,He:2019,Greenfeld:2019,Luz:2020,Eliasof:2020,Huang:2021,vanBetteray:2022}. These works differ in the details of their approaches, e.g., the choice of the loss function, the network architecture, and the kind of learning (supervised or unsupervised). The main difference to our work is that we have to address the gauge degrees of freedom. More precisely, our approach must be gauge-equivariant, i.e., the map implemented by the neural network must commute with local gauge transformations \cite{pmlr-v48-cohenc16,Cohen:2019}.

A number of papers have introduced  gauge-equivariant neural networks in the context of lattice quantum field theory: Refs.~\cite{Kanwar:2020xzo,Boyda:2020hsi,Abbott:2022zhs} mainly addressed the question of gauge-field sampling in several different theories, while Ref.~\cite{Favoni:2020reg} showed how any gauge-covariant function on the lattice can be approximated by neural networks. Our work builds on and extends these papers.

The structure of this paper is as follows.  In Sec.~\ref{sec:layers}, we introduce gauge-equivariant layers as the building blocks of the models we study in this work.  In Sec.~\ref{sec:wilson}, we discuss the problem of solving the preconditioned Dirac equation with the Wilson-clover Dirac operator.  In Sec.~\ref{sec:high}, we construct preconditioner models that address the high-mode component of the Dirac operator.  In Sec.~\ref{sec:low}, we discuss a model to address the low-mode component of the Dirac operator.  In Sec.~\ref{sec:multigrid}, we combine
the specialized models to a multi-grid model that addresses both the low-mode and high-mode components.  We conclude in Sec.~\ref{sec:summary}, where we also give an outlook to future work.

\section{Gauge-equivariant layers}\label{sec:layers}
In this section, we define the building blocks of the gauge-equivariant neural networks considered in this work and explain their properties in detail.
We begin with a discussion of the concepts of parallel transport and gauge equivariance.

\subsection{Parallel transport and gauge equivariance}
We consider a discrete $d$-dimensional space-time lattice with $L_\mu$ sites in dimension $\mu \in \{1,\ldots,d\}$ and $d \in \N$. The canonical unit vector in dimension $\mu$ is denoted by $\hat{\mu}$. The set of all lattice sites shall be $S = \{ (x_1,\ldots,x_{d}) \,\vert\, x_\mu \in \{1,\ldots,L_\mu\} \}$.
Consider a field $\phi:S \to V_I, x\mapsto \phi(x)$ with internal vector space $V_I$.  The internal vector space shall be a product of a gauge vector space $V_G  = \C^N$ and a non-gauge vector space $V_{\bar{G}}=\C^{\bar{N}}$ with $N,\bar{N} \in \N$ , i.e.,
\begin{align}
    V_I = V_G \otimes V_{\bar{G}} \,.
\end{align}
We also consider gauge fields $U_\mu : S \to \SU(N), x \mapsto U_\mu(x)$
with $\SU(N)$ acting on $V_G$.  The set of fields $\phi$ shall be ${\cal F}_\phi$, and the set of fields $U_\mu$ shall be ${\cal F}_U$.

We define the parallel-transport operator $T_p:{\cal F}_\phi \to {\cal F}_\phi, \phi \mapsto T_p \phi$ as
\begin{align}
     T_p &=  \Hop_{p_{n_p}} \cdots \Hop_{p_2} \Hop_{p_1}
\end{align}
for a path $p$ defined as the sequence $p_1,\ldots,p_{n_p}$ with $n_p \in \N$ and $p_i \in \{ \pm 1, \pm 2, \ldots, \pm d \}$.  The operator $\Hop_{p_i}:{\cal F}_\phi \to {\cal F}_\phi, \phi \mapsto \Hop_{p_i} \phi$ acts
on a field according to\footnote{Note that the operator $H_{p_i}$ does not act on the numerical value $\phi(x)$. Rather, it acts on the field $\phi$, resulting in the new field $H_{p_i}\phi$, which is then evaluated at $x$. Note also that in Eq.~\eqref{eq:Hop}, the information is transported from $x-\hat{p}_i$ to $x$.}
\begin{align}
  \label{eq:Hop}
  \Hop_{p_i} \phi(x) &= U^\dagger_{p_i}(x-\hat{p}_i) \phi(x-\hat{p}_i)
\end{align}
so as to transport information by a single hop in direction $\hat{p}_i$.  Here, we introduced the convention
$\hat{\nu} = - \hat{\mu}$ for $\nu=-\mu$, and we identify $U_{-\mu}(x) = U^\dagger_\mu(x - \hat{\mu})$. 
Addition and subtraction of coordinate tuples are defined component-wise.
Note that a single path $p$ defines the transport for any site $x \in S$ to
\begin{align}
 x^\prime = x + \sum_{i=1}^{n_p} \hat{p}_i
\end{align}
and may be illustrated using a representative starting point.  If $x^\prime = x$, the path is closed.  Note that the trivial path $0$ with $n_0=0$ and $T_0 = \1$ is allowed as well.

A field $\phi \in {\cal F}_\phi$ acquires a phase $\theta_\mu$ when translated by $L_\mu$ in direction $\hat{\mu}$, i.e.,
\begin{align}\label{eqn:phibc}
    \phi(x + L_\mu \hat{\mu}) = e^{i\theta_\mu}\phi(x)
\end{align}
for any coordinate tuple $x$.  A gauge field $U_\mu \in {\cal F}_U$ is periodic in all dimensions, i.e.,
\begin{align}
    U_\mu(x + L_\nu \hat{\nu}) = U_\mu(x)
\end{align}
with $\nu \in \{1,\ldots,d\}$.  These equations define $\phi(x)$ and $U_\mu(x)$ for all sites $x$ outside of $S$. 

\begin{figure}[tb]
    \centering
    \includegraphics[width=0.4\linewidth]{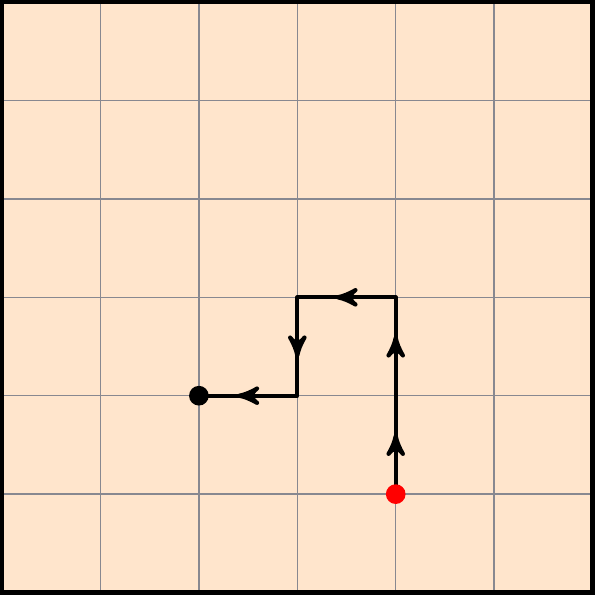}
    \caption{The path $p$ defining a parallel-transport operator $T_p$ can be visualized as a sequence of hops from a starting point (red) to an end point (black).}
    \label{fig:path}
\end{figure}

In Fig.~\ref{fig:path}, we illustrate the transport from the red starting point along a path $p$ to the black site.  This path corresponds to
\begin{align}
    T_p = \Hop_{-1} \Hop_{-2} \Hop_{-1} \Hop_{2} \Hop_{2}\,,
\end{align}
where $\hat{1}$ and $\hat{2}$ is the horizontal and vertical unit vector, respectively, in Fig.~\ref{fig:path}.

A gauge transformation is parametrized by a field $\Omega:S \to \SU(N),x\mapsto \Omega(x)$
that acts on all $\phi \in {\cal F}_\phi$ and $U_\mu \in {\cal F}_U$ by
\begin{align}
\phi(x) &\to \Omega(x) \phi(x) \,, \\
U_\mu(x) &\to \Omega(x) U_\mu(x) \Omega^\dagger(x+\hat{\mu}) \,.
\end{align}
It is straightforward to show that under such a gauge transformation we have
\begin{align}
 T_p \phi(x) \to \Omega(x) T_p \phi(x)
\end{align}
for any path $p$, i.e., the parallel-transport operator $T_p$ commutes with gauge transformations, and thus it is 
a gauge-equivariant operator. For a comprehensive discussion of gauge equivariance we refer to Ref.~\cite{Cohen:2019}.

\subsection{Parallel-transport convolutions}\label{sec:lptc}
The models discussed in this work will be composed of individual
layers that map $n$ input features $\phi_1,\ldots,\phi_n \in {\cal F}_\phi$ to $m$ output features $\psi_1,\ldots,\psi_m \in {\cal F}_\phi$.

We consider a parallel-transport convolution (PTC) layer defined by\footnote{Equation~\eqref{eq:PTC} is a convolution with kernel $W$ and input $\phi$, whose argument is shifted by $T_p$.}
\begin{align}
\label{eq:PTC}
 \psi_a(x) \stackrel{\rm PTC}{=} \sum_{b=1}^n \sum_{p \in P} W_a^{bp} T_p \phi_b(x)
\end{align}
for $a=1,\ldots,m$, with a set of paths $P$ and an endomorphism $W_a^{bp} \in \End(V_{\bar{G}})$.
This extends the definition of Ref.~\cite{Abbott:2022zhs} from nearest-neighbor hops to a sum over arbitrary paths.  For closed paths $p$, we recover the case discussed in Ref.~\cite{Favoni:2020reg}.  Note that in lattice QCD $W_a^{bp}$ is a $4\times4$ spin matrix.

We also consider a local parallel-transport convolution (LPTC) layer defined by
\begin{align}
\label{eq:LPTC}
 \psi_a(x) \stackrel{\rm LPTC}{=} \sum_{b=1}^n \sum_{p \in P} W_a^{bp}(x) T_p \phi_b(x)
\end{align}
with $W_a^{bp}:S \to \End(V_{\bar{G}}),x \mapsto W_a^{bp}(x)$.  Such a layer is also gauge equivariant and may be able to better address localized features. In the following we refer to the elements of $W$ as layer weights.

Since we intend to learn a \emph{linear} preconditioner in this work, we do not apply an activation function in these layers.  The expressivity of a deep network composed of such layers is therefore equivalent to a single layer with a larger set $P$.  Nevertheless, it may be computationally more efficient for a given problem to compose multiple layers with smaller sets $P$.

\begin{figure}[tb]
    \centering
    \includegraphics[width=.7\linewidth]{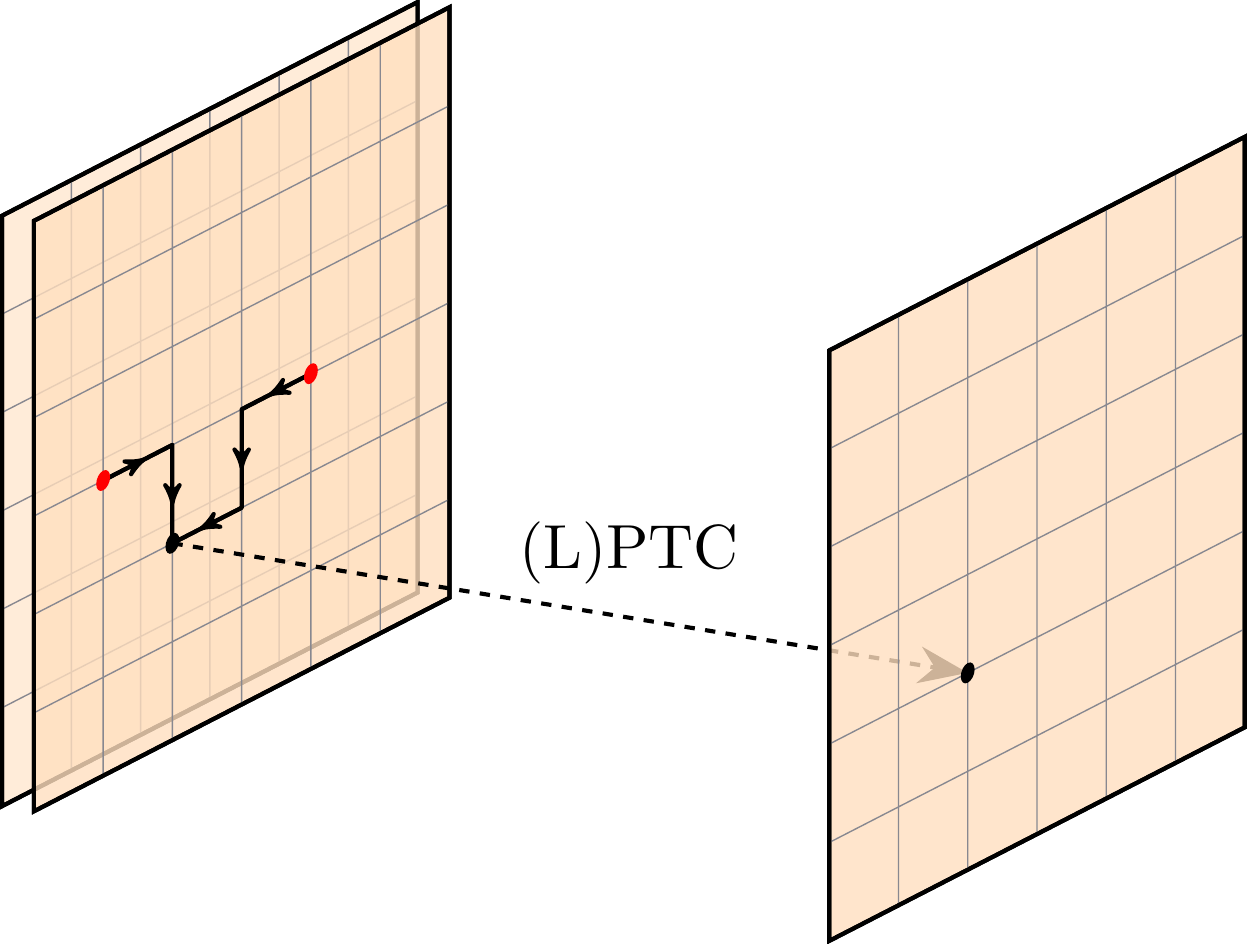}
    \caption{Graphical representation of a (L)PTC layer with two input features and one output feature. The planes represent the features. The layer is represented by the paths drawn and the dashed arrow.}
    \label{fig:ptc}
\end{figure}

In Fig.~\ref{fig:ptc}, we provide a graphical representation of a (L)PTC layer with two input features and one output feature and $P = \{ p_1, p_2 \}$ with
\begin{align}
    T_{p_1} = H_{-1} H_{-2} H_{-1} \,, \qquad
    T_{p_2} = H_{-2} H_{1} \,.
\end{align}

\subsection{Restriction and prolongation layers}\label{sec:restrictandprolong}
In order to let information propagate efficiently over long distances in terms of sites $x \in S$, we make use of the multi-grid paradigm \cite{Brannick:2007ue,Babich:2010qb}.  To this end, we consider a coarse grid
with lattice sites $\coarse{S}$ and a coarse field $\coarse{\phi}:\coarse{S} \to \coarse{V}_I,y \mapsto \coarse{\phi}(y)$ with coarse internal vector space $\coarse{V}_I$.  The set of such fields is denoted by ${\cal F}_{\coarse{\phi}}$.  Note that there are no gauge degrees of freedom in $\coarse{V}_I$.  

We define a restriction layer mapping a $\phi \in {\cal F}_\phi$ to a $\coarse{\psi} \in {\cal F}_{\coarse{\phi}}$ by
\begin{align}\label{eqn:restriction}
  \coarse{\psi}(y) \stackrel{\rm RL}{=} \sum_{x \in B(y)} W(y,x) \phi(x)
\end{align}
with $W:\coarse{S} \times S \to \Hom(V_I, \coarse{V}_I)$ and block map $B:\coarse{S} \to {\cal P}(S)$, where $\mathcal{P}$ denotes the power set. We also define a corresponding prolongation layer mapping a $\coarse{\phi} \in {\cal F}_{\coarse{\phi}}$ to a $\psi \in {\cal F}_\phi$ by
\begin{align}\label{eqn:prolongation}
  \psi(x) \stackrel{\rm PL}{=} W(y,x)^\dagger \coarse{\phi}(y)
\end{align}
for $x \in B(y)$.  In practice, we choose $B$ corresponding to a blocking in all dimensions.  The linear maps $W$ satisfy
\begin{align}
    \sum_{x \in B(y)} W(y,x) W(y,x)^\dagger =  \1_{\coarse{V}_I}\,,
\end{align}
where $\1_{\coarse{V}_I}$ is the identity in $\coarse{V}_I$.
These layers are straightforward to extend to the case of multiple input and output features.

\begin{figure}[tb]
    \centering
    \includegraphics[scale=0.33]{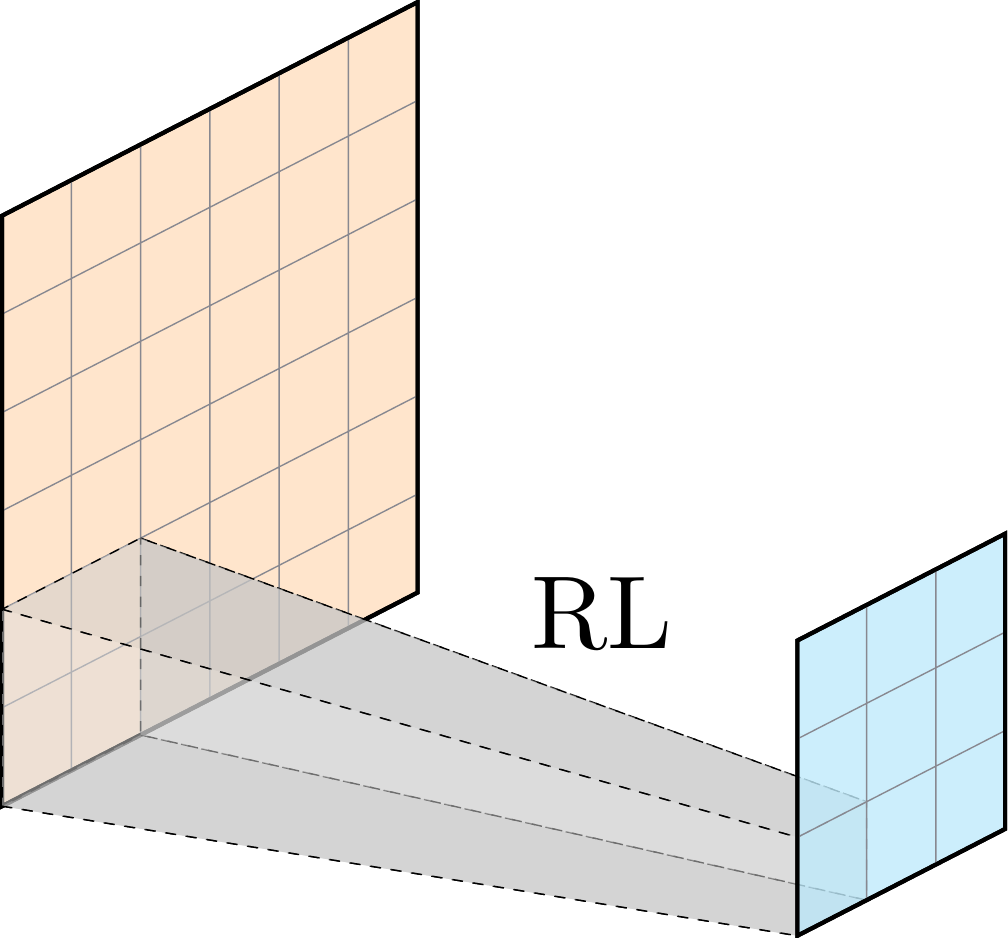}\hspace{1cm}\includegraphics[scale=0.33]{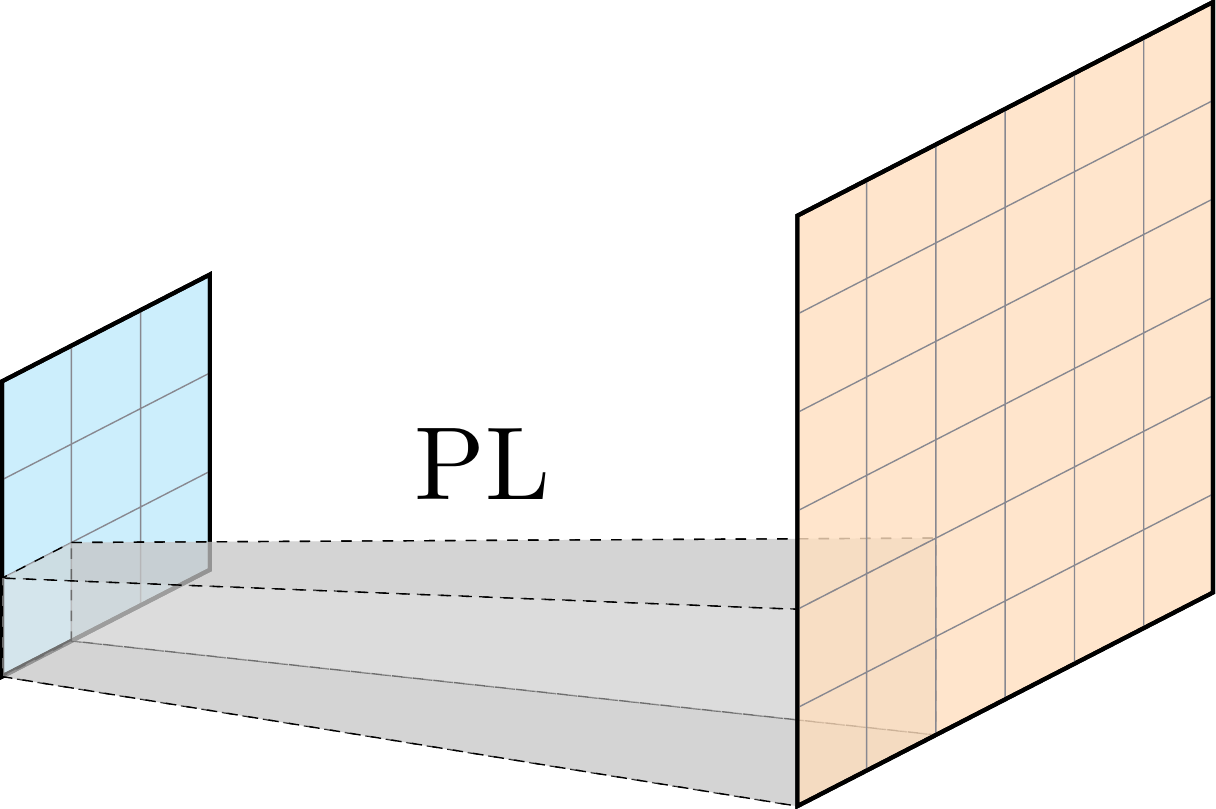}
    \caption{Graphical representation of the restriction layer (left) and prolongation layer (right) for a single feature. The layers are represented by the gray square frustums, while the input and output features are represented by the planes.}
    \label{fig:restrictprolong}
\end{figure}

The linear maps $W$ can be considered layer weights and are constructed from a list of vectors that are block-wise orthonormal, see Sec.~\ref{sec:low} for details.  The restriction and prolongation layers are gauge equivariant if
\begin{align}\label{eqn:coarsegaugetrafo}
W(y,x) \to W(y,x) \Omega(x)^\dagger
\end{align}
under a gauge transformation.
Note that since $\coarse{V}_I$ does not have gauge degrees of freedom there is no $\Omega(y)$ on the coarse grid.
We provide a graphical representation of the restriction and prolongation layers in Fig.~\ref{fig:restrictprolong}.

\subsection{Parallel and identity layers}
In this work, we consider models that act on a given input feature with multiple layers in parallel.  Consider applying a layer $L_i$ to
input features $\phi_1,\ldots,\phi_n$ mapping to output features $\psi_{i1},\ldots,\psi_{i{m_i}}$.  For several layers $L_1,\ldots,L_\ell$, we concatenate the output features $\psi_{11},\ldots,\psi_{1{m_1}},\ldots,\psi_{\ell1},\ldots,\psi_{\ell m_\ell}$.  The combination of layers $L_1,\ldots,L_\ell$ being applied in parallel can then be considered to be a single layer that maps features $\phi_1,\ldots,\phi_n$ to features $\psi_{11},\ldots,\psi_{1{m_1}},\ldots,\psi_{\ell1},\ldots,\psi_{\ell m_\ell}$.  

We also introduce an identity layer that maps the input features without modification to output features (which implies $m=n$). Such a layer is represented graphically by a single dashed arrow pointing from the input features to the output features. 

\begin{figure}[tb]
    \centering
    \includegraphics[width=\linewidth]{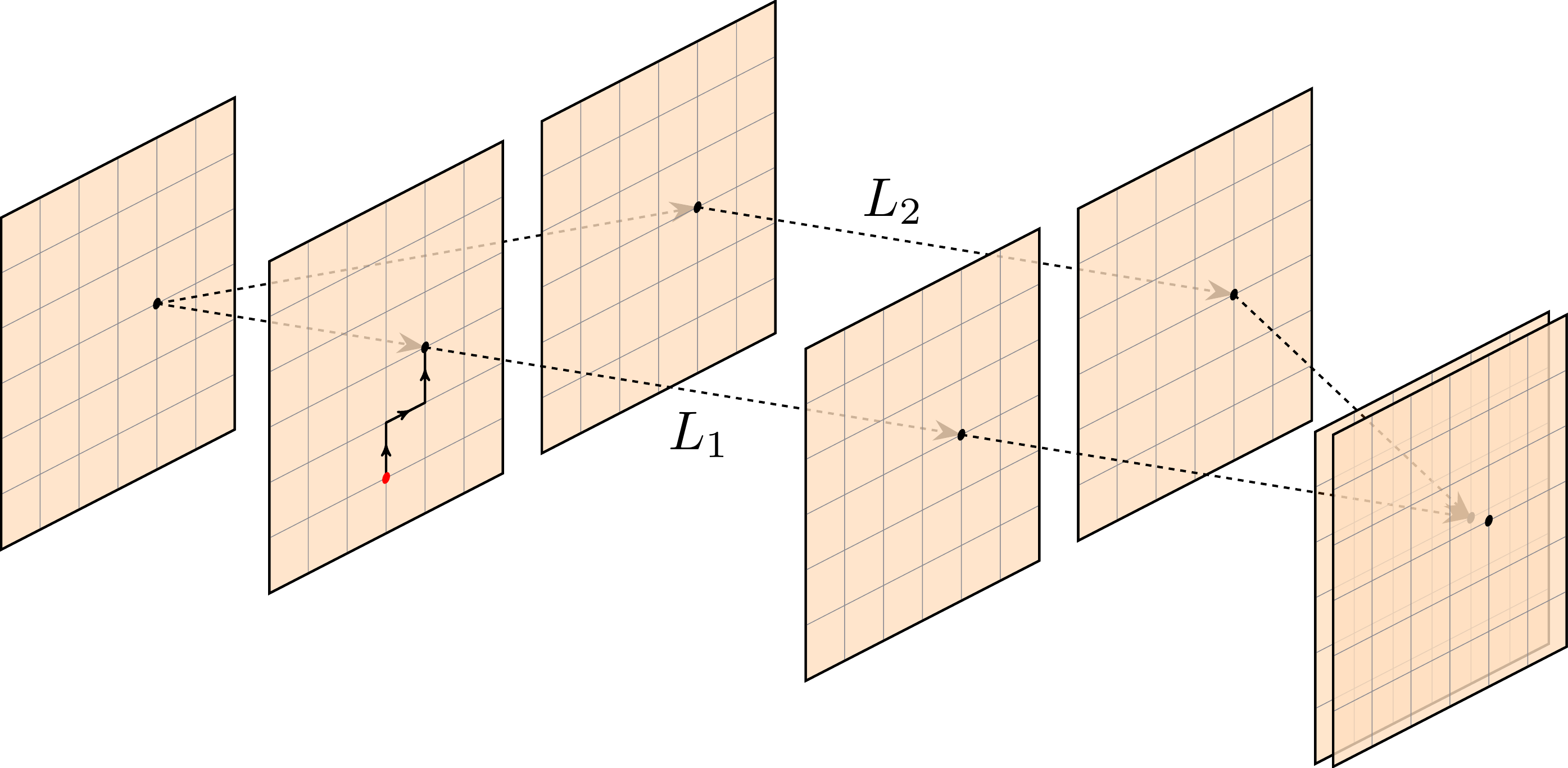}
    \caption{Graphical representation of two parallel layers $L_1$ and $L_2$ being applied to a single input feature and mapping to two output features. As before, the features are represented by planes. An identity layer (i.e., a copy operation) is represented by a dashed arrow. In this example, the only nontrivial layer is $L_1$, which includes a single path in \eqref{eq:PTC} or \eqref{eq:LPTC}.}
    \label{fig:parallel}
\end{figure}

We provide a graphical representation for the case of $n=1$, $\ell=2$, and $m_1=m_2=1$ in Fig.~\ref{fig:parallel}.

\subsection{Communication avoidance}\label{sec:commavoid}
In practice, the performance of a given model in terms of execution time
is crucial.  For problem sizes of interest to the lattice QCD community, a single problem will be distributed over multiple compute nodes that are connected by a communication network.  It is not uncommon that the time needed to exchange information between nodes exceeds the time each node spends performing floating-point operations.  Therefore it is an important paradigm in lattice QCD to investigate approaches that avoid communication between nodes even if it possibly increases the computational effort within a given node \cite{Luscher:2003qa,Osaki:2010vj,Babich:2011np,Tu:2021dvv}.
In this work, we also investigate layers which do not communicate between different sub-volumes that would typically be mapped to multiple nodes in an MPI job.  We perform such investigations by setting the gauge links $U_\mu$ that connect one such sub-volume to another to zero.  For such a modified model, we can then avoid the communication step between nodes altogether.

\section{The Wilson Dirac operator}\label{sec:wilson}
The main objective of this work is to precondition the Dirac
equation
\begin{align}\label{eqn:dirac}
    D u = b
\end{align}
with Dirac operator $D:{\cal F}_\phi \to {\cal F}_\phi$, source $b \in {\cal F}_\phi$, and solution $u \in {\cal F}_\phi$.  It is useful to interpret Eq.~\eqref{eqn:dirac} as a matrix equation with $u,b \in \C^k$ and invertible complex $k\times k$ matrix $D$
with
\begin{align}
  k = L_1 \cdots L_d  N  \bar{N} \,.
\end{align}
We train a model to play the role of an invertible complex $k\times k$ preconditioner matrix $M$
in
\begin{align}
  \label{eq:precond}
   (D M) M^{-1} u = b \,,
\end{align}
where we attempt to improve the condition number of $D M$ compared to $D$.
Ideally, $D M$ is close to the identity matrix up to a trivial scaling factor.  The Dirac matrix transforms as
\begin{align}
D \to \Omega D \Omega^{\dagger}
\end{align}
under a gauge transformation with block-diagonal matrix $\Omega=\oplus_{x\in S}\Omega(x)\otimes\1_{V_{\bar G}}$, which motivates the use of gauge-equivariant layers to construct $M$.

We first consider the Wilson Dirac operator \cite{Wilson:1974sk}
\begin{align}
    D_{\rm W} &= \frac12\sum_{\mu=1}^4 \gamma_\mu ( \Hop_{-\mu} - \Hop_{+\mu} ) + m \notag\\
    &\quad - \frac12 \sum_{\mu=1}^4 ( \Hop_{-\mu} + \Hop_{+\mu} - 2 )
\end{align}
with mass $m \in \R$ and Euclidean gamma matrices $\gamma_1,\ldots,\gamma_4$ satisfying the anti-commutation relation $ \gamma_\mu \gamma_\nu +\gamma_\nu \gamma_\mu=2\delta_{\mu\nu}$ with Kronecker delta $\delta_{\mu\nu}$.  This operator
can be mapped to a single PTC layer with a zero-hop path and eight one-hop paths.  

We add a clover term that includes closed paths consisting of four hops using
\begin{align}
    Q_{\mu\nu} &= \Hop_{-{\mu}}\Hop_{-{\nu}}\Hop_{+{\mu}}\Hop_{+{\nu}}  
    + \Hop_{-{\nu}}\Hop_{+{\mu}}\Hop_{+{\nu}}\Hop_{-{\mu}} \notag\\
& \quad 
+ \Hop_{+{\nu}} \Hop_{-{\mu}} \Hop_{-{\nu}} \Hop_{+{\mu}}   
+ \Hop_{+{\mu}}  \Hop_{+{\nu}} \Hop_{-{\mu}} \Hop_{-{\nu}}
\end{align}
to obtain the Wilson-clover Dirac operator \cite{Sheikholeslami:1985ij}
\begin{align}
D_{\rm WC} &= D_{\rm W} - \frac{c_{\rm sw}}{4} \sum_{\mu,\nu=1}^4 \sigma_{\mu\nu} F_{\mu\nu}
\end{align}
with $c_{\rm sw} \in \R$,
\begin{align}
 F_{\mu\nu} &= \frac18 (Q_{\mu\nu} - Q_{\nu\mu}) \,,
 \end{align}
 and
 \begin{align}
 \sigma_{\mu\nu} &= \frac12 (\gamma_\mu \gamma_\nu - \gamma_\nu \gamma_\mu) \,.
\end{align}
The operator $D_{\rm WC}$ can also be mapped to a single PTC layer, however, paths up to four hops are needed.  

\begin{figure}[tb]
    \centering
    \includegraphics{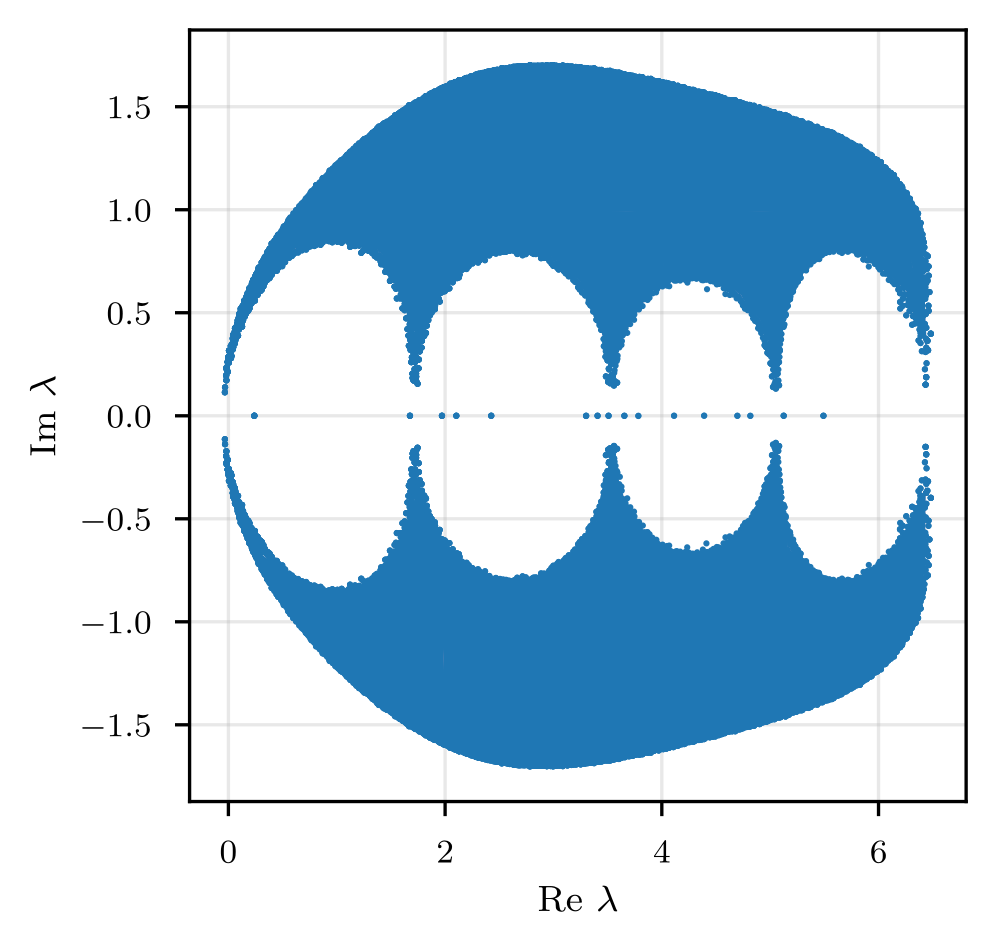}
    \caption{Eigenvalues $\lambda$ of the Wilson-clover Dirac operator with $m=-0.6$ and $c_{\rm sw}=1$ on a pure-Wilson-gauge configuration with $\beta=6$, $L_1=L_2=L_3=8$, and $L_4=16$.  The mass $m$ is tuned to near criticality for the experiments in this work.  We computed the boundaries of the spectrum using the Arnoldi method applied to $(D - \lambda)^{-1}$ for several carefully selected values of $\lambda$ and filled in the bulk of the spectrum by hand for illustrative purposes.}
    \label{fig:Dspectrum}
\end{figure}

For the numerical experiments presented in the following sections, we use gauge group $\SU(3)$ and the $D_{\rm WC}$ operator tuned to near criticality, i.e., the mass parameter is chosen such that the real part of the smallest eigenvalue is close to zero.  This provides a challenging problem even for the small lattice volume with $L_1=L_2=L_3=8$ and $L_4=16$ used in this work.  We set $m=-0.6$ and $c_{\rm sw}=1$
on a pure Wilson gauge configuration \cite{Wilson:1974sk} with coupling parameter $\beta=6.0$.
We use periodic boundary conditions also for the fields in ${\cal F}_\phi$, i.e., $\theta_\mu = 0$ in Eq.~\eqref{eqn:phibc}.
We show the spectrum of $D_{\rm WC}$ on a representative single gauge configuration in Fig.~\ref{fig:Dspectrum}.

We quantify the improvement achieved using the preconditioner $M$ by the reduction of the iteration count to solve Eq.~\eqref{eq:precond} to $10^{-8}$ precision in the preconditioned FGMRES \cite{FGMRES}.  We quote the iteration count gain defined as the iteration count of the unpreconditioned solve divided by the iteration count of the preconditioned solve.

The methods developed in this work also extend to other
Dirac matrices. However, particular challenges exist in some cases.
For example, in the case of domain-wall fermions \cite{Shamir:1993zy,Furman:1994ky}
the spectrum encircles the origin \cite{Brower:2020xmc,Boyle:2021wcf}, which limits the convergence of unpreconditioned solves of $D u = b$ using Krylov-subspace methods.

\section{High-mode preconditioners}\label{sec:high}

We want to learn a preconditioner $M$ that approximates $D^{-1}$. For this purpose it is useful to consider an eigendecomposition of $D$ and first construct optimal models for the high-mode and low-mode components separately.
We study the high-mode component in this section and the low-mode component in Sec.~\ref{sec:low}. We then combine the corresponding models in Sec.~\ref{sec:multigrid}.

\subsection{Model setup and training strategy}
\label{sec:highmodesetup}
The high-mode part of the spectrum of $D_{\rm WC}$ is related to the short-distance behavior. Therefore we expect a single layer with paths up to one hop to already show a gain in iteration count.  We consider a linear model $M$ mapping a vector $x$ to $M x$.
We employ a supervised learning approach and describe a single training step in the following.  

We first pick a random vector $v$ with components drawn from a Gaussian distribution with mean zero and unit standard deviation.  We then construct the cost function\footnote{Note that in Eq.~\eqref{eq:precond} we use $DM$, while in Eq.~\eqref{eq:cost} we use $MD$. If $DM$ is close to the identity, then so is $MD$, and thus Eq.~\eqref{eq:cost} is a suitable cost function.}
\begin{align}
\label{eq:cost}
 C = \vert M D_{\rm WC} v - v \vert^2
\end{align}
and its derivatives with respect to the model weights using backpropagation.  This corresponds to a batch of a single training tuple $(D_{\rm WC} v,v)$, where the model learns to map the first to the second component.  This cost function is dominated by the high modes of $D_{\rm WC}$ and is therefore similar in spirit to using the spectral radius \cite{Katrutsa:2017,Greenfeld:2019}.  Since we use a different random vector at every iteration our training data set is unbounded in size and there is no need to add a regulator.  This holds even for LPTC layers with a large number of model weights.
We then apply a single iteration of the Adam optimizer \cite{ADAM} with parameters $\beta_1=0.9$, $\beta_2=0.98$, and $\alpha=10^{-3}$ that gave good performance for the models considered in this work.  This process is repeated until the model weights are converged sufficiently.  

All layers and optimizers are implemented in the Grid Python Toolkit (GPT) \cite{GPT}, and corresponding code samples are provided in App.~\ref{app:code}.

\begin{figure}[tb]
    \centering
    \includegraphics{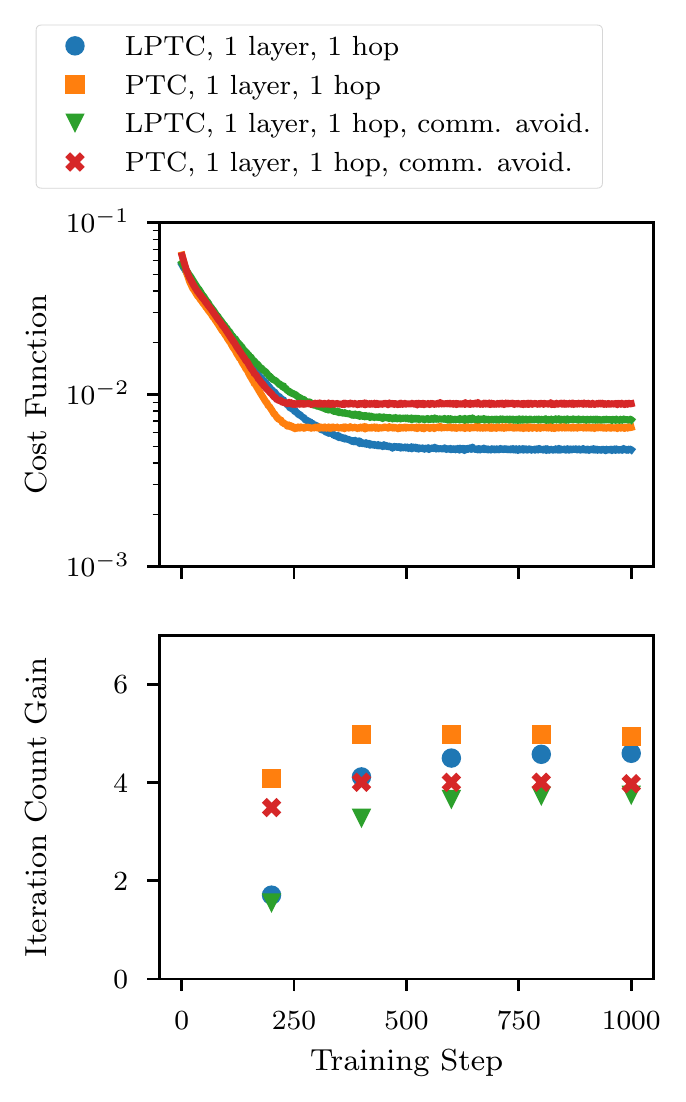}
    \caption{Convergence of the cost function \eqref{eq:cost} and iteration count gain for one-layer and one-hop high-mode preconditioners. The lattice volume is $8^3\times16$, and the local volume for the communication-avoiding version is $4^3\times8$.}
    \label{fig:high-mode-one-layer-one-hop}
\end{figure}

\subsection{Locality and communication avoidance}
In Fig.~\ref{fig:high-mode-one-layer-one-hop} we compare the performance of 
single-layer models with a maximum of one hop.  They correspond to a version of Fig.~\ref{fig:ptc} with a single input and output feature and nine paths corresponding to
\begin{align}\label{eqn:onehoppaths}
 T_0 &= \1 \,, \notag\\
 T_1 &= H_{{1}} \,, & T_2 &= H_{{2}} \,, \notag\\
 T_3 &= H_{{3}} \,, & T_4 &= H_{{4}} \,, \notag\\
 T_5 &= H_{-{1}} \,, & T_6 &= H_{-{2}} \,, \notag\\
 T_7 &= H_{-{3}} \,, & T_8 &= H_{-{4}} \,.
\end{align}
We also investigate communication-avoiding versions with local volume $4^3 \times 8$.  We find that the LPTC models do not perform better in terms of iteration count gain than the PTC models. However, the LPTC models require more training compared to the PTC models.  The slower convergence is expected due to the much larger number weights in the LPTC models.  We find that eliminating communication between sub-volumes,
as described in Sec.~\ref{sec:commavoid}, only leads to a modest reduction in performance.  After translating the iteration count gain to a reduction in time-to-solution, we may therefore find the communication-avoiding models to perform best.

\begin{figure}[tb]
    \centering
    \includegraphics{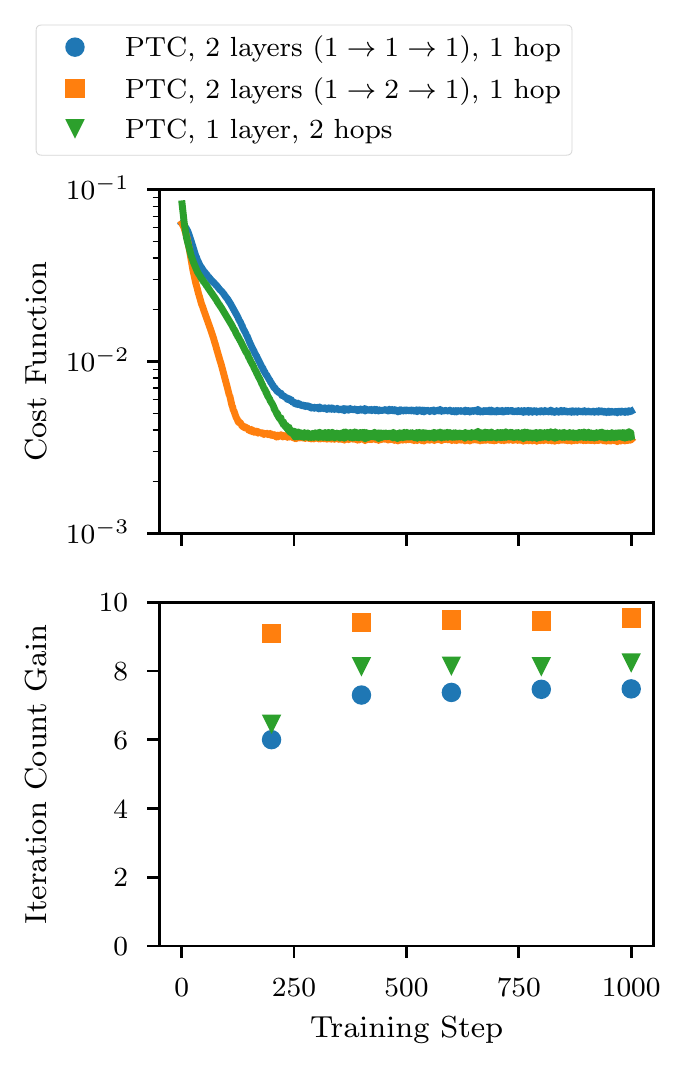}
    \caption{Convergence of the cost function \eqref{eq:cost} and iteration count gain for two-layer and two-hop high-mode preconditioners.}
    \label{fig:high-mode-two-layer-two-hop}
\end{figure}

\subsection{Multiple hops and deep networks}
In Fig.~\ref{fig:high-mode-two-layer-two-hop}, we investigate models with multiple hops either in a single layer or distributed over two layers.  We use one-hop layers
with paths defined in Eq.~\eqref{eqn:onehoppaths} as well as a two-hop layer
extending this set by all combinations
\begin{align}
 H_{a} H_{b}
\end{align}
for $a,b \in \{ -4,-3,-2,-1,1,2,3,4 \}$ with $a \neq - b$.  The two-hop layer therefore has 65 distinct paths compared to the 9 paths of the one-hop layer.

The first model that we investigate stacks two one-hop layers with one input and one output feature back-to-back.  We denote this model as ``2 layers ($1\to 1 \to 1$), 1 hop.''  The second model is similar but has two output features in the first layer and correspondingly two input features in the second layer.  We denote this model as ``2 layers ($1 \to 2 \to 1$), 1 hop.''  The third model consists of a single two-hop layer as described above.

We find that the second model performs best and gives approximately twice the iteration count gain of the corresponding single-layer models with a maximum of one hop shown in Fig.~\ref{fig:high-mode-one-layer-one-hop}.  Since the layers are linear, the two-layer models are not more expressive compared to the single-layer model with two hops.  We therefore expect the third model to be able to match the performance of the second model with a sufficiently improved training procedure.  It is not surprising that the second model can be trained more efficiently compared to the third model given that it has a smaller number of weights.  We conclude that while deep models do not increase expressivity, the computational effort needed to train deep models may be reduced compared to a corresponding shallow model with more paths.

\begin{figure}[tb]
    \centering
    \includegraphics{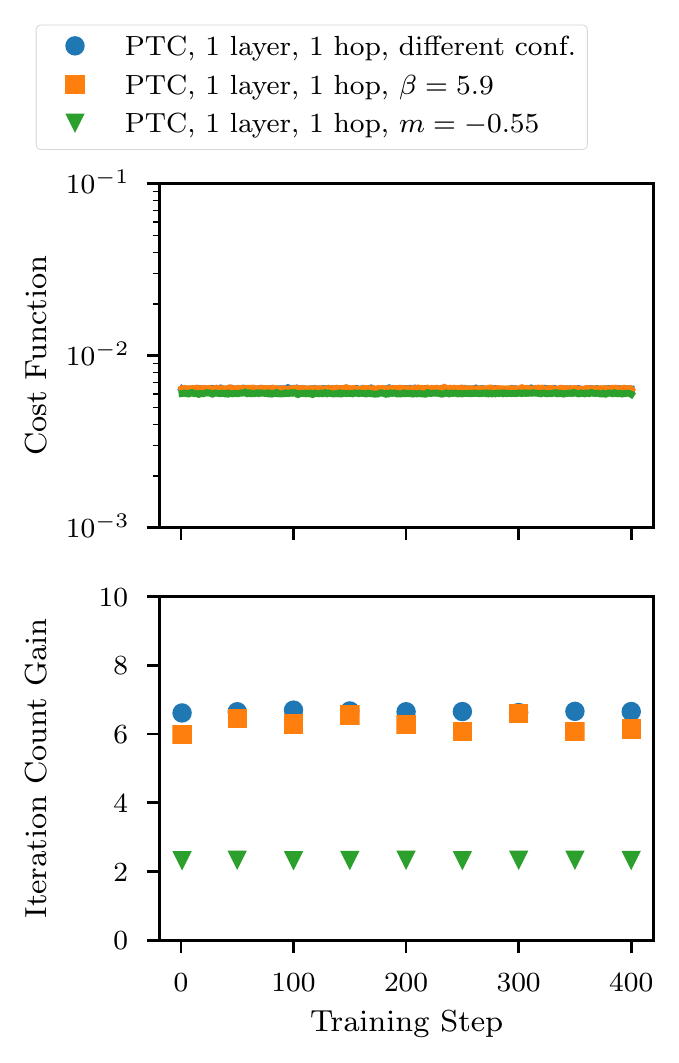}
    \caption{Convergence of the cost function \eqref{eq:cost} and iteration count gain for one-layer and one-hop high-mode preconditioners.  We re-train the model of Fig.~\ref{fig:high-mode-one-layer-one-hop} for a different gauge configuration in the same ensemble, for a different value of $\beta=5.9$, and for a different mass value of $m=-0.55$.  The network performs well in all cases even without re-training.}
    \label{fig:high-mode-retraining}
\end{figure}

\subsection{Transfer learning}
In Fig.~\ref{fig:high-mode-retraining} we investigate how well the one-layer one-hop PTC model of Fig.~\ref{fig:high-mode-one-layer-one-hop} that was trained on a given gauge configuration with $\beta=6.0$ and $m=-0.6$ performs when it is used in the case of (i) a different gauge configuration of the same gauge ensemble, (ii) a gauge configuration of a different ensemble with $\beta=5.9$, and (iii) the same gauge configuration but with a different mass $m=-0.55$.  In all cases, we investigate the performance without re-training and after additional re-training steps following the same procedure as for the initial training.  We find that the high-mode preconditioner model does not require re-training to efficiently perform in all three cases.  Once such a model is trained, it can be used efficiently for different gauge configurations of the same and similar ensembles.  We note that the maximum iteration count gain for mass $m=-0.55$ is significantly reduced.  In this case, however, the spectrum is not well tuned to criticality and the initial problem is therefore less challenging.
Comparing with Fig.~\ref{fig:high-mode-one-layer-one-hop}, we also observe a modest fluctuation in iteration count gain between different configurations.

\section{Low-mode preconditioners}\label{sec:low}

We now turn to the low-mode component in the eigendecomposition of $D$. Since the low-mode component corresponds to the long-distance behavior of the Dirac operator $D$, it is not efficient to use the layers discussed in Sec.~\ref{sec:high} since a rather deep network composed of such layers would be needed to propagate information over sufficiently long distances.  
The multi-grid paradigm, however, is ideally suited to address this issue.  In this section, we focus solely on the low-mode component and then combine low modes and high modes in Sec.~\ref{sec:multigrid}.

\subsection{Model setup and training strategy}
In the multi-grid approach, we define an additional coarser version of the lattice
as well as restriction and prolongation operations that map between the fine and coarse lattices.  These operations must preserve the low-mode component of $D$ \cite{Luscher:2007se}. 

To achieve this, we first find vectors $u_1,\ldots,u_s$ in the near-null space of $D$, i.e., vectors that satisfy
\begin{align}\label{eqn:nearnullproblem}
  D u_i \approx 0 
\end{align}
with null vector 0 and $i\in \{1,\ldots,s\}$ for $s=\dim(\coarse{V}_I)$.  These vectors are then blocked such that one site $y \in \coarse{S}$ on the coarse lattice corresponds to a set of sites, or block, $B(y) \subset S$ on the fine lattice.  Let us denote such a blocked vector, which lives on the sites $B(y)$, by $u_i^y$.  One then defines an inner product within each block $B(y)$ and orthonormalizes the vectors $u_1^y,\ldots,u_s^y$ within each block according to this inner product.  The resulting vectors are labeled $\bar{u}_1^y,\ldots,\bar{u}_s^y$.  The linear map $W^\dagger$ discussed in Sec.~\ref{sec:restrictandprolong} is then defined as
\begin{align}
 W(y,x)^\dagger = \sum_{i=1}^s \bar{u}_i^y(x) \hat{e}_i^\dagger
\end{align}
with standard basis $\hat{e}_1,\ldots,\hat{e}_s$ of $\coarse{V}_I$ and $x \in B(y)$.

In practice a good approximation of such vectors $u_i$ can be found by applying the FGMRES solver for matrix $D$ with source vector 0 and a random vector as initial guess.  This procedure removes high-mode components in $u_i$, leaving a linear combination of low-modes. We follow this approach in the numerical experiments presented in the following. 
 While high precision is not needed, we solve to $10^{-8}$ precision to avoid an additional tuning step.  We use a coarse grid of size $2^3 \times 4$ and a list of 12 near-null vectors $u_1,\ldots,u_{12}$.  

We define a coarse-grid operator
\begin{align}
 \coarse{D} = R D_{\rm WC} P
\end{align}
with restriction matrix $R$ and prolongation matrix $P$ that are defined according to Eqs.~\eqref{eqn:restriction} and \eqref{eqn:prolongation}.  We then train a coarse-grid model $\coarse{M}$ that contains a single LPTC layer with gauge fields $U_\mu = \1$, $V_G=\C^1$, $V_{\bar{G}} = \tilde{V}_I$, and use only zero-hop and one-hop paths corresponding to $\{ H_1, H_2, H_3, H_4, H_{-4} \}$.  We omit the $H_{-1}$, $H_{-2}$, and $H_{-3}$ paths since they are redundant on a $2^3 \times 4$ coarse grid with periodic boundary conditions.  The gauge fields are replaced with the identity since the coarse fields do not have a gauge degree of freedom.  We refer to this special case of the LPTC layer as cLPTC in the following.

We follow the training procedure described in Sec.~\ref{sec:highmodesetup} but replace the cost
function with
\begin{align}
\label{eq:costcoarse}
 C = \vert \coarse{M} \coarse{D} v - v \vert^2 \,.
\end{align}
It is worth noting that one could have considered a different cost function
\begin{align}
 C^\prime = \vert \coarse{M} v - \coarse{D}^{-1} v \vert^2
\end{align}
in order to project more strongly on the low modes of $\coarse{D}$.  In this case, however, the training tuples require the somewhat costly inversion of $\coarse{D}$.  We find that the cost function Eq.~\eqref{eq:costcoarse} is sufficient for the purpose of training the coarse-grid model.  This point will be revisited when we train a combined multi-grid model in Sec.~\ref{sec:multigrid}.

Note that the gauge equivariance of the restriction and prolongation layers is guaranteed if every vector $u_i$ is a linear combination of eigenmodes of $D$ with gauge-invariant coefficients.  In our procedure the coefficients are gauge invariant in the statistical average over random initial guess vectors.
Furthermore, note that the weights $W$ of the restriction and prolongation layers could also be learned directly \cite{Katrutsa:2017,Greenfeld:2019}.  We leave the systematic study of learning the restriction and prolongation layers, including explicitly gauge-equivariant versions, to future work.

\subsection{Results}\label{sec:resultslow}
\begin{figure}[tb]
    \centering
    \includegraphics{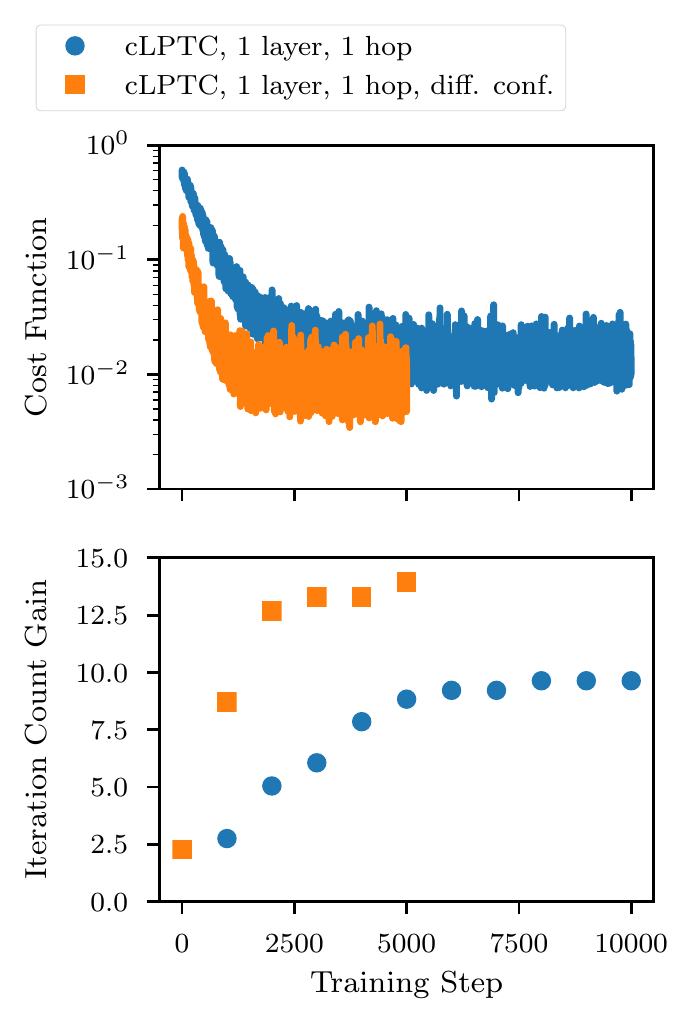}
    \caption{Convergence of the cost function \eqref{eq:costcoarse} and iteration count gain for one-layer and one-hop low-mode preconditioners.   We show both the initial training in blue as well as the performance of the trained model on a different gauge field of the same gauge ensemble in orange.  We find that after a moderate amount of re-training, the model performs well on a different gauge configuration.}
    \label{fig:low-mode-retraining}
\end{figure}
In Fig.~\ref{fig:low-mode-retraining}, we show the cost function \eqref{eq:costcoarse} and the iteration count gain for the training of the coarse-grid model $\coarse{M}$.  In this case, we consider the iteration count gain for the inverse of $\coarse{D}$.  We find that a significantly longer training process is needed compared to the high-mode preconditioner models of Sec.~\ref{sec:high}.

We also investigate using the fully trained model from a given gauge configuration and applying it to a different gauge configuration.  We use the same definition of the restriction and prolongation layers on the different gauge configuration to preserve the definition of $\tilde{D}$.  For the same reason we also use the same seeds for the random number generator to generate the initial guess for the fields $u_{1},\ldots,u_{12}$.  We find that after a modest amount of re-training the model performs very well on the different gauge configuration.  The re-training phase is significantly shorter compared to the initial training phase.  We note that the maximum iteration count gain again differs to some degree between configurations.

\section{Multi-grid preconditioners}\label{sec:multigrid}

In the previous sections we successfully trained separate models $M$ to approximate the short-distance and long-distance features of $D^{-1}$.
In this section we combine them to obtain a model that approximates $D^{-1}$ over a wide range of distances.

\subsection{Smoother model setup and training strategy}\label{sec:smoother}
We first create a version of the short-distance model that accepts a second input feature, which provides an initial guess.  This model plays the role of a smoother in the multi-grid paradigm.  The initial guess is provided by the long-distance model acting on the coarse grid.

Concretely, we aim to find a sequence of $u_k$ that approximately solve $D u = b$ such that the equation becomes exact in the $k \to \infty$ limit.  The smoother then maps the tuple $(u_k,b)$ to $u_{k+1}$.
If we have a high-mode model $M_{\rm h}$ that approximates $D^{-1}$ sufficiently well this can be achieved by the iterative relaxation approach
\begin{align}\label{eqn:relax}
 u_{k+1} &= (\1 - M_{\rm h} D) u_k + M_{\rm h} b \notag \\
 &= u_k + M_{\rm h} (b - D u_k) \,.
\end{align}
This approach is also commonly referred to as defect correction with defect $b- D u_k$.

Since both $D$ and the high-mode model $M_{\rm h}$ can be represented by (L)PTC layers we should be able to train a model $M_{\rm s}$ only composed of (L)PTC layers to map $(u_k,b)$ to a $u_{k+r}$ for $r \in \N^+$.  Such a model has two input features and one output feature.  We may construct $M_{\rm s}$ using $2r$ (L)PTC layers stacked back-to-back since each iteration of Eq.~\eqref{eqn:relax} corresponds to two (L)PTC layers.  All but the final layer need two output features.

\begin{figure}[tb]
    \centering
    \includegraphics{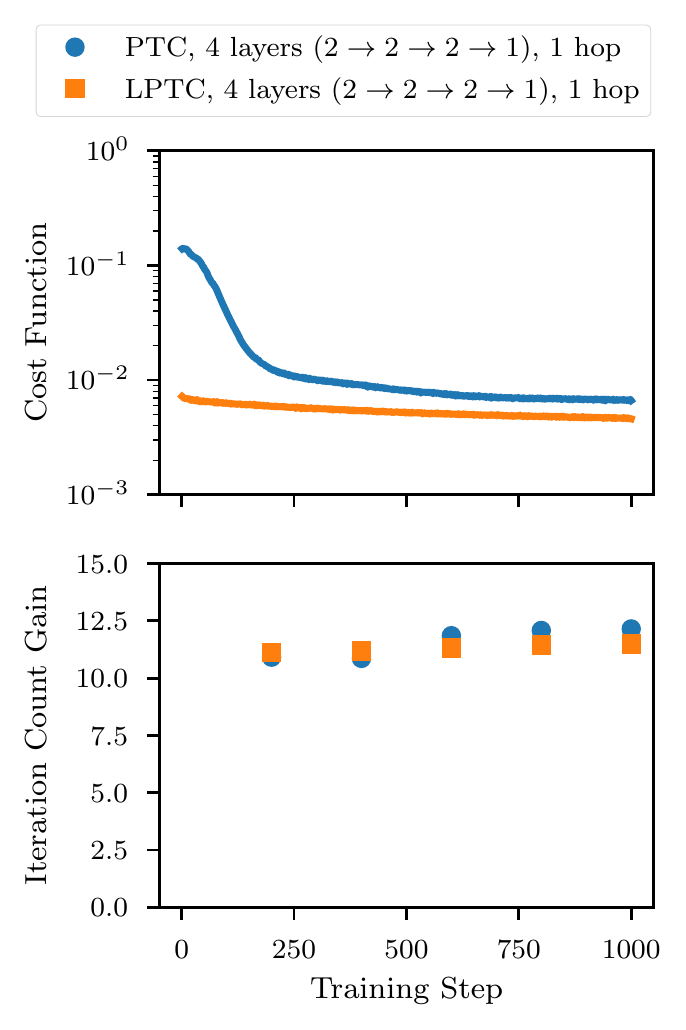}
    \caption{Convergence of the cost function \eqref{eq:costsmoother} and iteration count gain for four-layer and one-hop smoother.  The iteration count gain is studied for the case of zero initial guess.  We first train the PTC model and use the result as initial weights for the LPTC model.}
    \label{fig:smoother}
\end{figure}

In order to choose a reasonable value for $r$, we studied the performance of the final multi-grid preconditioner described below and found that $r=2$ performed significantly better than $r=1$.  We therefore train the model $M_{\rm s}$ for $r=2$ using the cost
function
\begin{align}
\label{eq:costsmoother}
    C = \vert M_{\rm s}(u_k,b) - u_{k+r} \vert^2
\end{align}
with random vectors $(u_k,b)$ and $u_{k+r}$ given by Eq.~\eqref{eqn:relax}.  We use the same optimizer as in Secs.~\ref{sec:high} and \ref{sec:low}.

In Fig.~\ref{fig:smoother}, we show the training progress.  The iteration count gain is obtained by using $M_{\rm s}$ with initial guess zero as a preconditioner for $D u = b$.  We use both PTC and LPTC layers with zero-hop and one-hop paths.  We expect these models to yield an iteration count gain of approximately twice the iteration count gain of the corresponding high-mode models shown in Fig.~\ref{fig:high-mode-one-layer-one-hop} because of $r=2$.  We find that this expectation is satisfied by our data. 
In Fig.~\ref{fig:smoother}, we first train the PTC model and then use the model weights as initial values for the LPTC model (using the same value for every site $x$).  We find no additional benefit by using the LPTC model.

\subsection{Multi-grid model setup and training strategy}
\begin{figure*}[tb]
    \centering
    \includegraphics[width=\linewidth]{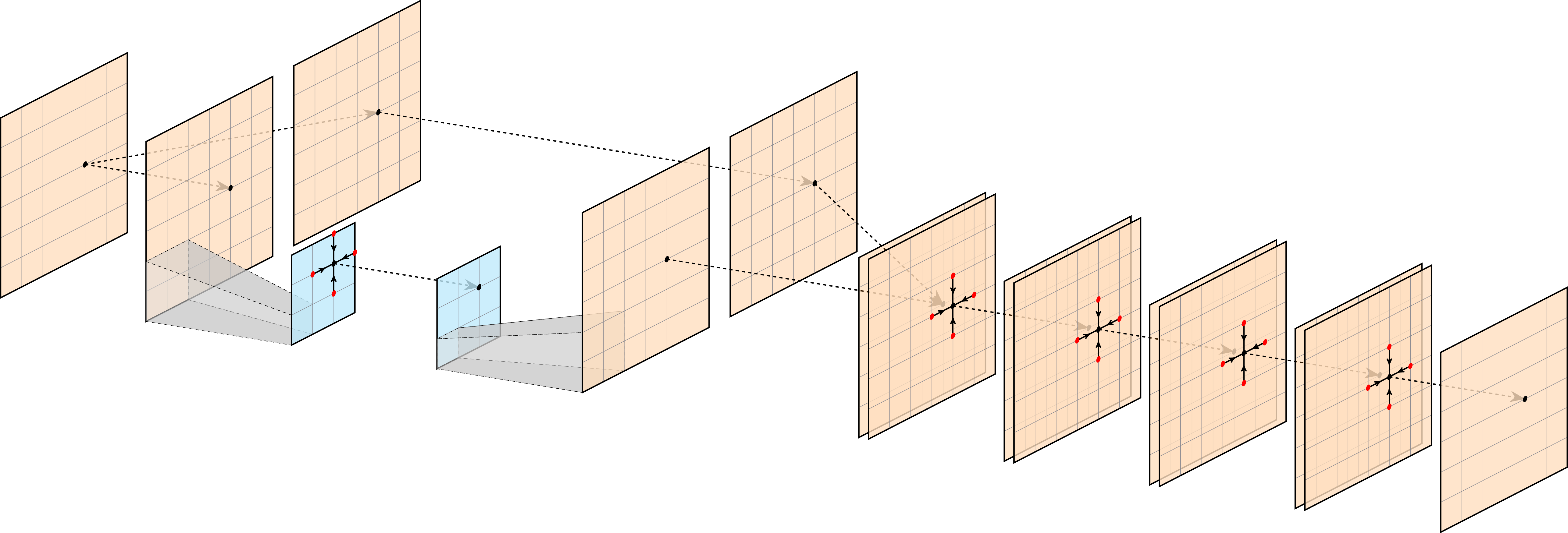}
    \caption{The combined two-level multi-grid model studied in this work.  The use of the multi-grid paradigm allows for the efficient transport of information over both short and long-distances.  Additional levels can be introduced by recursively replacing the coarse-grid layer (limited by the blue features) by the entire model as presented above.}
    \label{fig:multigridmodel}
\end{figure*}

We are now ready to combine the individual models to a complete multi-grid model $M$ as shown in Fig.~\ref{fig:multigridmodel}.  We start by duplicating the input feature.  One copy is preserved for the smoother, while the other copy is restricted to the coarse grid, where we apply the coarse-grid model of Sec.~\ref{sec:low}.  The result is then prolonged to the fine grid, and both the copy of the initial feature and the result of the coarse-grid model are combined to two input features for the last four layers.  These layers are the smoother that we have learned in Sec.~\ref{sec:smoother}.

We may expect this combined model to work well by using the weights obtained in the training of the respective model components.  The model performance may, however, be further improved by continued training of the complete multi-grid model $M$.
For such additional training, we need to modify the cost function of Secs.~\ref{sec:low} and \ref{sec:high} such that both the low-mode and high-mode components of $D$ constrain the model in the training phase.  To this end, we use
\begin{align}\label{eq:costmg}
 C = \vert M b_h - u_h \vert^2 + \vert M b_\ell - u_\ell \vert^2
\end{align}
with $b_h = D_{\rm WC} v_1$, $u_h=v_1$, $b_\ell = v_2$, and $u_\ell = D_{\rm WC}^{-1} v_2$.  Here, $v_1$ and $v_2$ are random vectors normalized such that $\vert b_h \vert = \vert b_\ell \vert = 1$.  We therefore use a batch size of two with one training tuple geared towards the high-mode component and the other training tuple geared towards the low-mode component of $D_{\rm WC}$.  We can shift the focus of the training between both components by adding a relative weight factor to Eq.~\eqref{eq:costmg}.

\begin{figure}[!b]
    \centering
    \includegraphics{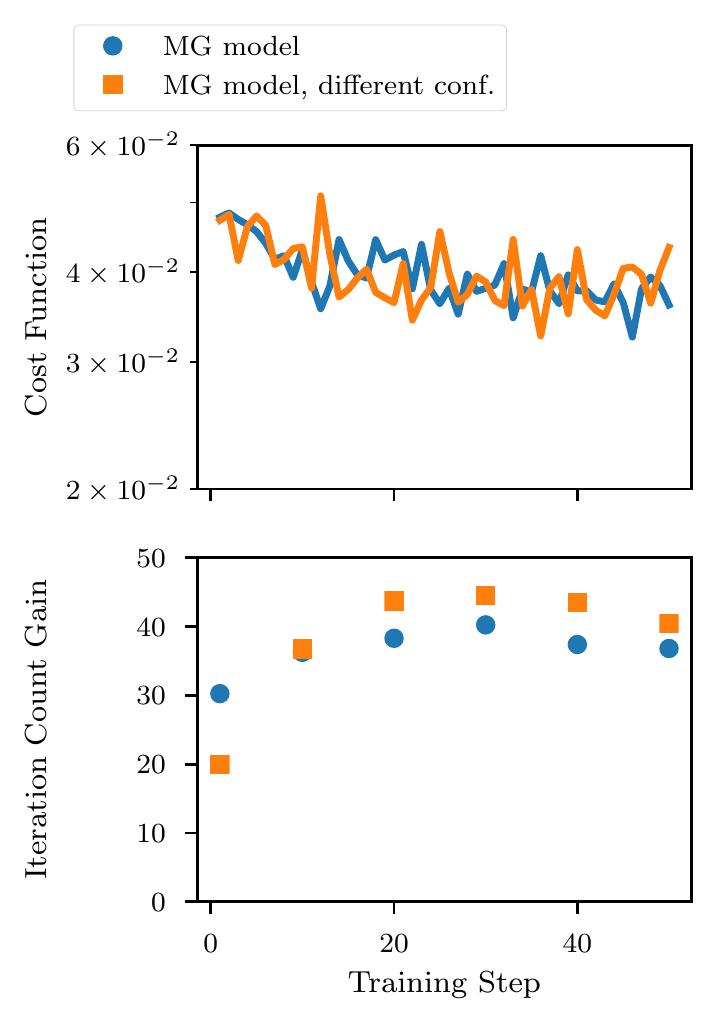}
    \caption{Convergence of the cost function \eqref{eq:costmg} and iteration count gain for the complete multi-grid model.  We use the weights of the individually trained model components as starting point and show further improvement by training the combined model.  The model also performs well on a different gauge configuration and quickly converges to optimum performance after a modest amount of re-training. 
}
    \label{fig:multigrid}
\end{figure}

\subsection{Results}

In Fig.~\ref{fig:multigrid}, we show the performance of the multi-grid (MG) model with initial weights taken from the trained model components as well as progress achieved by continued training of the combined model $M$.  From the start, the model performs substantially better than the smoother by itself. Continued training of the combined model further improves the iteration count gain to approximately 40.  Such continued training converges within the first 20 training steps.

We also study using the multi-grid model trained on one configuration applied to a different gauge configuration of the same gauge ensemble.  In Fig.~\ref{fig:multigrid}, we show that after a brief re-training phase of only 20 training steps, the model performs optimally on the different gauge configuration as well.

Note that for concreteness we only present results for a two-level multi-grid preconditioner in this work.  The extension to multiple levels is straightforward.  In Fig.~\ref{fig:multigridmodel}, one merely has to replace the coarse-grid layer limited by the blue features by the entire model as presented in Fig.~\ref{fig:multigridmodel}.  By repeating this process $n$ times, one obtains an $(n+2)$--level multi-grid preconditioner.

Also note that we use a rather small lattice volume of $8^3 \times 16$ in this work.  In future work, we will investigate multi-grid models in more challenging large-volume simulations, where even larger iteration count gains should be achievable.

\section{Summary and outlook}\label{sec:summary}

In this paper we have initiated a program to use gauge-equivariant neural networks to learn preconditioners in lattice QCD. 
We introduced a number of building blocks from which suitable models can be constructed: (i) parallel-transport convolution layers that can include arbitrary paths, with either global or local weights, (ii) restriction and prolongation layers that implement the multi-grid paradigm, and (iii) parallel layers that act on a single input feature.

To solve the Dirac equation for the Wilson-clover Dirac operator we have first constructed models that approximate the high-mode and low-mode component of the operator separately.  We then combined these models in a two-level multi-grid model, which can be extended straightforwardly to an arbitrary number of levels. In all cases we found that the models reduce the iteration count of the outer solver significantly, e.g., by up to $O(40)$ in the multi-grid model. We also found that transfer learning works: If we consider another gauge configuration (for the same or a slightly different value of $\beta$) or a slightly different quark mass, only a modest amount of re-training (or none at all) is required for the model to perform efficiently again.

We also introduced a communication-avoiding algorithm in which layers do not transfer information between sub-volumes assigned to different MPI processes. In our numerical experiments we found that the performance, i.e., the iteration count gain, of the corresponding model is only slightly reduced. We expect that on large supercomputers, the wall-clock time saved by avoiding communication more than compensates for this modest reduction.

There are many interesting directions which we plan to explore in future work. For example, we will attempt to learn the weights $W$ of the restriction and prolongation layers directly, without computing the near-null vectors explicitly. Also, we will investigate the space of possible models that can be constructed from our building blocks in a more comprehensive manner. Furthermore, we plan to perform benchmarks that measure the cost of (re-) training and applying our models and compare the overall wall-clock time to standard state-of-the-art multi-grid methods. It would also be worthwhile to apply our ideas to Dirac operators whose spectrum encircles the origin, such as in the case of domain-wall fermions. Finally, our finding that very little, if any, re-training is needed between configurations suggests that the present approach could also be beneficial in the generation of gauge-field configurations by Markov chain Monte Carlo.

\appendix

\section{GPT code listings}\label{app:code}
In this appendix, we provide Grid Python Toolkit (GPT) \cite{GPT}
code listings to implement the models used in this work.  We first import the library and load a gauge field $U$:
\begin{lstlisting}[language=Python]
import gpt as g

# load gauge field
U = g.load("gauge_field")
grid = U[0].grid
\end{lstlisting}
The layer drawn in Fig.~\ref{fig:ptc} corresponds to
\begin{lstlisting}[language=Python]
# object types for QCD
ot_i = g.ot_vector_spin_color(4,3)
ot_w = g.ot_matrix_spin(4)

# two distinct paths
paths = [
    g.path().f(0).f(1).f(0),
    g.path().f(1).b(0)
]

# define an abbreviation
l = g.ml.layer

# define the layer of Fig. 2
fig2 = layer.parallel_transport_convolution(
    grid, U, paths, ot_i, ot_w, 2, 1
)

\end{lstlisting}
in the case of lattice QCD.
Next, we define restriction and prolongation layers to a coarse grid of size $4^4$ defined using vectors $\bar{u}_i$ as
\begin{lstlisting}[language=Python]
# define coarse grid
coarse_grid = g.grid([4,4,4,4], g.double)

# load \bar{u}_i vectors
u_bar = g.load("u_bar")

# create blocking map
b = g.block.map(coarse_grid, basis)

# create restriction and prolongation layers
restrict = l.block.project(b)
prolong = l.block.promote(b)
\end{lstlisting}
Note that in the numerical work in this paper, we used a $2^3 \times 4$ coarse grid, while we present the $4^4$ case here since it lifts the degeneracy of paths mentioned in Sec.~\ref{sec:low}.

The complete multi-grid preconditioner model of Fig.~\ref{fig:multigridmodel} corresponds to
\begin{lstlisting}[language=Python]
# define abbreviations
lptc = l.local_parallel_transport_convolution
ptc = l.parallel_transport_convolution

# identies on coarse grid
one=g.complex(coarse_grid)
one[:]=1

I=[g.copy(one) for i in range(4)]

# coarse-grid vector space
cot_i = g.ot_vector_complex_additive_group(
    len(u_bar)
)
cot_w = g.ot_matrix_complex_additive_group(
    len(u_bar)
)

# consider only nearest-neighbor hops
paths = [
    g.path().forward(i) 
    for i in range(4)
] + [
    g.path().backward(i) 
    for i in range(4)
]

# coarse-grid layer
def coarse_lptc(n_in, n_out):
    return lptc(
        coarse_grid, I, paths, 
        cot_i, cot_w, n_in, n_out
    )

# fine-grid layer
def fine_ptc(n_in, n_out):
    return ptc(
        grid, U, paths, ot_i,
        ot_w, n_in, n_out
    )

# combined multi-grid model
model_multi_grid = g.ml.model.sequence(
    l.parallel(
        l.sequence(),
        l.sequence(
            restrict,
            coarse_lptc(1, 1),
            prolong
        )
    ),
    fine_ptc(2, 2),
    fine_ptc(2, 2),
    fine_ptc(2, 2),
    fine_ptc(2, 1),
)
\end{lstlisting}

\clearpage

\bibliography{references}

\begin{thebibliography}{35}%
\makeatletter
\providecommand \@ifxundefined [1]{%
 \@ifx{#1\undefined}
}%
\providecommand \@ifnum [1]{%
 \ifnum #1\expandafter \@firstoftwo
 \else \expandafter \@secondoftwo
 \fi
}%
\providecommand \@ifx [1]{%
 \ifx #1\expandafter \@firstoftwo
 \else \expandafter \@secondoftwo
 \fi
}%
\providecommand \natexlab [1]{#1}%
\providecommand \enquote  [1]{``#1''}%
\providecommand \bibnamefont  [1]{#1}%
\providecommand \bibfnamefont [1]{#1}%
\providecommand \citenamefont [1]{#1}%
\providecommand \href@noop [0]{\@secondoftwo}%
\providecommand \href [0]{\begingroup \@sanitize@url \@href}%
\providecommand \@href[1]{\@@startlink{#1}\@@href}%
\providecommand \@@href[1]{\endgroup#1\@@endlink}%
\providecommand \@sanitize@url [0]{\catcode `\\12\catcode `\$12\catcode
  `\&12\catcode `\#12\catcode `\^12\catcode `\_12\catcode `\%12\relax}%
\providecommand \@@startlink[1]{}%
\providecommand \@@endlink[0]{}%
\providecommand \url  [0]{\begingroup\@sanitize@url \@url }%
\providecommand \@url [1]{\endgroup\@href {#1}{\urlprefix }}%
\providecommand \urlprefix  [0]{URL }%
\providecommand \Eprint [0]{\href }%
\providecommand \doibase [0]{https://doi.org/}%
\providecommand \selectlanguage [0]{\@gobble}%
\providecommand \bibinfo  [0]{\@secondoftwo}%
\providecommand \bibfield  [0]{\@secondoftwo}%
\providecommand \translation [1]{[#1]}%
\providecommand \BibitemOpen [0]{}%
\providecommand \bibitemStop [0]{}%
\providecommand \bibitemNoStop [0]{.\EOS\space}%
\providecommand \EOS [0]{\spacefactor3000\relax}%
\providecommand \BibitemShut  [1]{\csname bibitem#1\endcsname}%
\let\auto@bib@innerbib\@empty
\bibitem [{\citenamefont {Kronfeld}\ \emph {et~al.}(2022)\citenamefont
  {Kronfeld} \emph {et~al.}}]{USQCD:2022mmc}%
  \BibitemOpen
  \bibfield  {author} {\bibinfo {author} {\bibfnamefont {A.~S.}\ \bibnamefont
  {Kronfeld}} \emph {et~al.} (\bibinfo {collaboration} {USQCD}),\ }\href@noop
  {} {\bibinfo {title} {{Lattice QCD and Particle Physics}}} (\bibinfo {year}
  {2022}),\ \Eprint {https://arxiv.org/abs/2207.07641} {arXiv:2207.07641
  [hep-lat]} \BibitemShut {NoStop}%
\bibitem [{\citenamefont {Boyle}\ \emph {et~al.}(2022)\citenamefont {Boyle}
  \emph {et~al.}}]{Boyle:2022ncb}%
  \BibitemOpen
  \bibfield  {author} {\bibinfo {author} {\bibfnamefont {P.}~\bibnamefont
  {Boyle}} \emph {et~al.},\ }\href@noop {} {\bibinfo {title} {{Lattice QCD and
  the Computational Frontier}}} (\bibinfo {year} {2022}),\ \Eprint
  {https://arxiv.org/abs/2204.00039} {arXiv:2204.00039 [hep-lat]} \BibitemShut
  {NoStop}%
\bibitem [{\citenamefont {{C. Lehner et al.}}()}]{GPT}%
  \BibitemOpen
  \bibfield  {author} {\bibinfo {author} {\bibnamefont {{C. Lehner et al.}}},\
  }\href {https://github.com/lehner/gpt} {\bibinfo {title} {{Grid Python
  Toolkit (GPT)}}}\BibitemShut {NoStop}%
\bibitem [{\citenamefont {Brannick}\ \emph {et~al.}(2008)\citenamefont
  {Brannick}, \citenamefont {Brower}, \citenamefont {Clark}, \citenamefont
  {Osborn},\ and\ \citenamefont {Rebbi}}]{Brannick:2007ue}%
  \BibitemOpen
  \bibfield  {author} {\bibinfo {author} {\bibfnamefont {J.}~\bibnamefont
  {Brannick}}, \bibinfo {author} {\bibfnamefont {R.~C.}\ \bibnamefont
  {Brower}}, \bibinfo {author} {\bibfnamefont {M.~A.}\ \bibnamefont {Clark}},
  \bibinfo {author} {\bibfnamefont {J.~C.}\ \bibnamefont {Osborn}},\ and\
  \bibinfo {author} {\bibfnamefont {C.}~\bibnamefont {Rebbi}},\ }\bibfield
  {title} {\bibinfo {title} {{Adaptive Multigrid Algorithm for Lattice QCD}},\
  }\href {https://doi.org/10.1103/PhysRevLett.100.041601} {\bibfield  {journal}
  {\bibinfo  {journal} {Phys. Rev. Lett.}\ }\textbf {\bibinfo {volume} {100}},\
  \bibinfo {pages} {041601} (\bibinfo {year} {2008})},\ \Eprint
  {https://arxiv.org/abs/0707.4018} {arXiv:0707.4018 [hep-lat]} \BibitemShut
  {NoStop}%
\bibitem [{\citenamefont {Babich}\ \emph {et~al.}(2010)\citenamefont {Babich},
  \citenamefont {Brannick}, \citenamefont {Brower}, \citenamefont {Clark},
  \citenamefont {Manteuffel}, \citenamefont {McCormick}, \citenamefont
  {Osborn},\ and\ \citenamefont {Rebbi}}]{Babich:2010qb}%
  \BibitemOpen
  \bibfield  {author} {\bibinfo {author} {\bibfnamefont {R.}~\bibnamefont
  {Babich}}, \bibinfo {author} {\bibfnamefont {J.}~\bibnamefont {Brannick}},
  \bibinfo {author} {\bibfnamefont {R.~C.}\ \bibnamefont {Brower}}, \bibinfo
  {author} {\bibfnamefont {M.~A.}\ \bibnamefont {Clark}}, \bibinfo {author}
  {\bibfnamefont {T.~A.}\ \bibnamefont {Manteuffel}}, \bibinfo {author}
  {\bibfnamefont {S.~F.}\ \bibnamefont {McCormick}}, \bibinfo {author}
  {\bibfnamefont {J.~C.}\ \bibnamefont {Osborn}},\ and\ \bibinfo {author}
  {\bibfnamefont {C.}~\bibnamefont {Rebbi}},\ }\bibfield  {title} {\bibinfo
  {title} {{Adaptive multigrid algorithm for the lattice Wilson-Dirac
  operator}},\ }\href {https://doi.org/10.1103/PhysRevLett.105.201602}
  {\bibfield  {journal} {\bibinfo  {journal} {Phys. Rev. Lett.}\ }\textbf
  {\bibinfo {volume} {105}},\ \bibinfo {pages} {201602} (\bibinfo {year}
  {2010})},\ \Eprint {https://arxiv.org/abs/1005.3043} {arXiv:1005.3043
  [hep-lat]} \BibitemShut {NoStop}%
\bibitem [{\citenamefont {Frommer}\ \emph {et~al.}(2014)\citenamefont
  {Frommer}, \citenamefont {Kahl}, \citenamefont {Krieg}, \citenamefont
  {Leder},\ and\ \citenamefont {Rottmann}}]{Frommer:2013fsa}%
  \BibitemOpen
  \bibfield  {author} {\bibinfo {author} {\bibfnamefont {A.}~\bibnamefont
  {Frommer}}, \bibinfo {author} {\bibfnamefont {K.}~\bibnamefont {Kahl}},
  \bibinfo {author} {\bibfnamefont {S.}~\bibnamefont {Krieg}}, \bibinfo
  {author} {\bibfnamefont {B.}~\bibnamefont {Leder}},\ and\ \bibinfo {author}
  {\bibfnamefont {M.}~\bibnamefont {Rottmann}},\ }\bibfield  {title} {\bibinfo
  {title} {{Adaptive Aggregation Based Domain Decomposition Multigrid for the
  Lattice Wilson Dirac Operator}},\ }\href {https://doi.org/10.1137/130919507}
  {\bibfield  {journal} {\bibinfo  {journal} {SIAM J. Sci. Comput.}\ }\textbf
  {\bibinfo {volume} {36}},\ \bibinfo {pages} {A1581} (\bibinfo {year}
  {2014})},\ \Eprint {https://arxiv.org/abs/1303.1377} {arXiv:1303.1377
  [hep-lat]} \BibitemShut {NoStop}%
\bibitem [{\citenamefont {Brannick}\ \emph {et~al.}(2016)\citenamefont
  {Brannick}, \citenamefont {Frommer}, \citenamefont {Kahl}, \citenamefont
  {Leder}, \citenamefont {Rottmann},\ and\ \citenamefont
  {Strebel}}]{Brannick:2014vda}%
  \BibitemOpen
  \bibfield  {author} {\bibinfo {author} {\bibfnamefont {J.}~\bibnamefont
  {Brannick}}, \bibinfo {author} {\bibfnamefont {A.}~\bibnamefont {Frommer}},
  \bibinfo {author} {\bibfnamefont {K.}~\bibnamefont {Kahl}}, \bibinfo {author}
  {\bibfnamefont {B.}~\bibnamefont {Leder}}, \bibinfo {author} {\bibfnamefont
  {M.}~\bibnamefont {Rottmann}},\ and\ \bibinfo {author} {\bibfnamefont
  {A.}~\bibnamefont {Strebel}},\ }\bibfield  {title} {\bibinfo {title}
  {{Multigrid Preconditioning for the Overlap Operator in Lattice QCD}},\
  }\href {https://doi.org/10.1007/s00211-015-0725-6} {\bibfield  {journal}
  {\bibinfo  {journal} {Numer. Math.}\ }\textbf {\bibinfo {volume} {132}},\
  \bibinfo {pages} {463} (\bibinfo {year} {2016})},\ \Eprint
  {https://arxiv.org/abs/1410.7170} {arXiv:1410.7170 [hep-lat]} \BibitemShut
  {NoStop}%
\bibitem [{\citenamefont {Brower}\ \emph {et~al.}(2018)\citenamefont {Brower},
  \citenamefont {Clark}, \citenamefont {Strelchenko},\ and\ \citenamefont
  {Weinberg}}]{Brower:2018ymy}%
  \BibitemOpen
  \bibfield  {author} {\bibinfo {author} {\bibfnamefont {R.~C.}\ \bibnamefont
  {Brower}}, \bibinfo {author} {\bibfnamefont {M.~A.}\ \bibnamefont {Clark}},
  \bibinfo {author} {\bibfnamefont {A.}~\bibnamefont {Strelchenko}},\ and\
  \bibinfo {author} {\bibfnamefont {E.}~\bibnamefont {Weinberg}},\ }\bibfield
  {title} {\bibinfo {title} {{Multigrid algorithm for staggered lattice
  fermions}},\ }\href {https://doi.org/10.1103/PhysRevD.97.114513} {\bibfield
  {journal} {\bibinfo  {journal} {Phys. Rev. D}\ }\textbf {\bibinfo {volume}
  {97}},\ \bibinfo {pages} {114513} (\bibinfo {year} {2018})},\ \Eprint
  {https://arxiv.org/abs/1801.07823} {arXiv:1801.07823 [hep-lat]} \BibitemShut
  {NoStop}%
\bibitem [{\citenamefont {{R.~C.~Brower, M.~A.~Clark, D.~Howarth,
  E.~S.~Weinberg}}(2020)}]{Brower:2020xmc}%
  \BibitemOpen
  \bibfield  {author} {\bibinfo {author} {\bibnamefont {{R.~C.~Brower,
  M.~A.~Clark, D.~Howarth, E.~S.~Weinberg}}},\ }\bibfield  {title} {\bibinfo
  {title} {{Multigrid for chiral lattice fermions: Domain wall}},\ }\href
  {https://doi.org/10.1103/PhysRevD.102.094517} {\bibfield  {journal} {\bibinfo
   {journal} {Phys. Rev. D}\ }\textbf {\bibinfo {volume} {102}},\ \bibinfo
  {pages} {094517} (\bibinfo {year} {2020})},\ \Eprint
  {https://arxiv.org/abs/2004.07732} {arXiv:2004.07732 [hep-lat]} \BibitemShut
  {NoStop}%
\bibitem [{\citenamefont {Boyle}\ and\ \citenamefont
  {Yamaguchi}(2021)}]{Boyle:2021wcf}%
  \BibitemOpen
  \bibfield  {author} {\bibinfo {author} {\bibfnamefont {P.}~\bibnamefont
  {Boyle}}\ and\ \bibinfo {author} {\bibfnamefont {A.}~\bibnamefont
  {Yamaguchi}},\ }\href@noop {} {\bibinfo {title} {{Comparison of Domain Wall
  Fermion Multigrid Methods}}} (\bibinfo {year} {2021}),\ \Eprint
  {https://arxiv.org/abs/2103.05034} {arXiv:2103.05034 [hep-lat]} \BibitemShut
  {NoStop}%
\bibitem [{\citenamefont {Trottenberg}\ \emph {et~al.}(2000)\citenamefont
  {Trottenberg}, \citenamefont {Oosterlee},\ and\ \citenamefont
  {Schuller}}]{trottenberg2000multigrid}%
  \BibitemOpen
  \bibfield  {author} {\bibinfo {author} {\bibfnamefont {U.}~\bibnamefont
  {Trottenberg}}, \bibinfo {author} {\bibfnamefont {C.}~\bibnamefont
  {Oosterlee}},\ and\ \bibinfo {author} {\bibfnamefont {A.}~\bibnamefont
  {Schuller}},\ }\href {https://books.google.de/books?id=9ysyNPZoR24C} {\emph
  {\bibinfo {title} {Multigrid}}}\ (\bibinfo  {publisher} {Elsevier Science},\
  \bibinfo {year} {2000})\BibitemShut {NoStop}%
\bibitem [{\citenamefont {Katrutsa}\ \emph {et~al.}(2017)\citenamefont
  {Katrutsa}, \citenamefont {Daulbaev},\ and\ \citenamefont
  {Oseledets}}]{Katrutsa:2017}%
  \BibitemOpen
  \bibfield  {author} {\bibinfo {author} {\bibfnamefont {A.}~\bibnamefont
  {Katrutsa}}, \bibinfo {author} {\bibfnamefont {T.}~\bibnamefont {Daulbaev}},\
  and\ \bibinfo {author} {\bibfnamefont {I.}~\bibnamefont {Oseledets}},\
  }\href@noop {} {\bibinfo {title} {{Deep Multigrid: learning prolongation and
  restriction matrices}}} (\bibinfo {year} {2017}),\ \Eprint
  {https://arxiv.org/abs/1711.03825} {arXiv:1711.03825 [math.NA]} \BibitemShut
  {NoStop}%
\bibitem [{\citenamefont {He}\ and\ \citenamefont {Xu}(2019)}]{He:2019}%
  \BibitemOpen
  \bibfield  {author} {\bibinfo {author} {\bibfnamefont {J.}~\bibnamefont
  {He}}\ and\ \bibinfo {author} {\bibfnamefont {J.}~\bibnamefont {Xu}},\
  }\bibfield  {title} {\bibinfo {title} {{MgNet}: A unified framework of
  multigrid and convolutional neural network},\ }\href
  {https://doi.org/10.1007/s11425-019-9547-2} {\bibfield  {journal} {\bibinfo
  {journal} {Sci. China Math.}\ }\textbf {\bibinfo {volume} {62}},\ \bibinfo
  {pages} {1331} (\bibinfo {year} {2019})},\ \Eprint
  {https://arxiv.org/abs/1901.10415} {arXiv:1901.10415 [cs.CV]} \BibitemShut
  {NoStop}%
\bibitem [{\citenamefont {Greenfeld}\ \emph {et~al.}(2019)\citenamefont
  {Greenfeld}, \citenamefont {Galun}, \citenamefont {Basri}, \citenamefont
  {Yavneh},\ and\ \citenamefont {Kimmel}}]{Greenfeld:2019}%
  \BibitemOpen
  \bibfield  {author} {\bibinfo {author} {\bibfnamefont {D.}~\bibnamefont
  {Greenfeld}}, \bibinfo {author} {\bibfnamefont {M.}~\bibnamefont {Galun}},
  \bibinfo {author} {\bibfnamefont {R.}~\bibnamefont {Basri}}, \bibinfo
  {author} {\bibfnamefont {I.}~\bibnamefont {Yavneh}},\ and\ \bibinfo {author}
  {\bibfnamefont {R.}~\bibnamefont {Kimmel}},\ }\bibfield  {title} {\bibinfo
  {title} {{Learning to Optimize Multigrid {PDE} Solvers}},\ }in\ \href
  {https://proceedings.mlr.press/v97/greenfeld19a.html} {\emph {\bibinfo
  {booktitle} {Proceedings of the 36th International Conference on Machine
  Learning}}}\ (\bibinfo {year} {2019})\ pp.\ \bibinfo {pages} {2415--2423},\
  \Eprint {https://arxiv.org/abs/1902.10248} {arXiv:1902.10248 [math.NA]}
  \BibitemShut {NoStop}%
\bibitem [{\citenamefont {Luz}\ \emph {et~al.}(2020)\citenamefont {Luz},
  \citenamefont {Galun}, \citenamefont {Maron}, \citenamefont {Basri},\ and\
  \citenamefont {Yavneh}}]{Luz:2020}%
  \BibitemOpen
  \bibfield  {author} {\bibinfo {author} {\bibfnamefont {I.}~\bibnamefont
  {Luz}}, \bibinfo {author} {\bibfnamefont {M.}~\bibnamefont {Galun}}, \bibinfo
  {author} {\bibfnamefont {H.}~\bibnamefont {Maron}}, \bibinfo {author}
  {\bibfnamefont {R.}~\bibnamefont {Basri}},\ and\ \bibinfo {author}
  {\bibfnamefont {I.}~\bibnamefont {Yavneh}},\ }\bibfield  {title} {\bibinfo
  {title} {Learning algebraic multigrid using graph neural networks},\ }in\
  \href {https://dl.acm.org/doi/10.5555/3524938.3525540} {\emph {\bibinfo
  {booktitle} {Proceedings of the 37th International Conference on Machine
  Learning}}}\ (\bibinfo {year} {2020})\ pp.\ \bibinfo {pages} {6489--6499},\
  \Eprint {https://arxiv.org/abs/2003.05744} {arXiv:2003.05744 [cs.LG]}
  \BibitemShut {NoStop}%
\bibitem [{\citenamefont {Eliasof}\ \emph {et~al.}(2020)\citenamefont
  {Eliasof}, \citenamefont {Ephrath}, \citenamefont {Ruthotto},\ and\
  \citenamefont {Treister}}]{Eliasof:2020}%
  \BibitemOpen
  \bibfield  {author} {\bibinfo {author} {\bibfnamefont {M.}~\bibnamefont
  {Eliasof}}, \bibinfo {author} {\bibfnamefont {J.}~\bibnamefont {Ephrath}},
  \bibinfo {author} {\bibfnamefont {L.}~\bibnamefont {Ruthotto}},\ and\
  \bibinfo {author} {\bibfnamefont {E.}~\bibnamefont {Treister}},\ }\href@noop
  {} {\bibinfo {title} {{MGIC: Multigrid-in-Channels Neural Network
  Architectures}}} (\bibinfo {year} {2020}),\ \Eprint
  {https://arxiv.org/abs/2011.09128} {arXiv:2011.09128 [cs.CV]} \BibitemShut
  {NoStop}%
\bibitem [{\citenamefont {Huang}\ \emph {et~al.}(2021)\citenamefont {Huang},
  \citenamefont {Li},\ and\ \citenamefont {Xi}}]{Huang:2021}%
  \BibitemOpen
  \bibfield  {author} {\bibinfo {author} {\bibfnamefont {R.}~\bibnamefont
  {Huang}}, \bibinfo {author} {\bibfnamefont {R.}~\bibnamefont {Li}},\ and\
  \bibinfo {author} {\bibfnamefont {Y.}~\bibnamefont {Xi}},\ }\bibfield
  {title} {\bibinfo {title} {Learning optimal multigrid smoothers via neural
  networks},\ }\href {https://doi.org/10.1137/21M1430030} {\bibfield  {journal}
  {\bibinfo  {journal} {SIAM J. Sci. Comput.}\ ,\ \bibinfo {pages} {S199}}
  (\bibinfo {year} {2021})},\ \Eprint {https://arxiv.org/abs/2102.12071}
  {arXiv:2102.12071 [math.NA]} \BibitemShut {NoStop}%
\bibitem [{\citenamefont {van Betteray}\ \emph {et~al.}(2022)\citenamefont {van
  Betteray}, \citenamefont {Rottmann},\ and\ \citenamefont
  {Kahl}}]{vanBetteray:2022}%
  \BibitemOpen
  \bibfield  {author} {\bibinfo {author} {\bibfnamefont {A.}~\bibnamefont {van
  Betteray}}, \bibinfo {author} {\bibfnamefont {M.}~\bibnamefont {Rottmann}},\
  and\ \bibinfo {author} {\bibfnamefont {K.}~\bibnamefont {Kahl}},\ }\href@noop
  {} {\bibinfo {title} {{MGiaD: Multigrid in all dimensions. Efficiency and
  robustness by coarsening in resolution and channel dimensions}}} (\bibinfo
  {year} {2022}),\ \Eprint {https://arxiv.org/abs/2211.05525} {arXiv:2211.05525
  [cs.CV]} \BibitemShut {NoStop}%
\bibitem [{\citenamefont {Cohen}\ and\ \citenamefont
  {Welling}(2016)}]{pmlr-v48-cohenc16}%
  \BibitemOpen
  \bibfield  {author} {\bibinfo {author} {\bibfnamefont {T.}~\bibnamefont
  {Cohen}}\ and\ \bibinfo {author} {\bibfnamefont {M.}~\bibnamefont
  {Welling}},\ }\bibfield  {title} {\bibinfo {title} {{Group Equivariant
  Convolutional Networks}},\ }in\ \href
  {https://dl.acm.org/doi/10.5555/3045390.3045705} {\emph {\bibinfo {booktitle}
  {Proceedings of The 33rd International Conference on Machine Learning}}}\
  (\bibinfo {year} {2016})\ pp.\ \bibinfo {pages} {2990--2999},\ \Eprint
  {https://arxiv.org/abs/1602.07576} {arXiv:1602.07576 [cs.LG]} \BibitemShut
  {NoStop}%
\bibitem [{\citenamefont {Cohen}\ \emph {et~al.}(2019)\citenamefont {Cohen},
  \citenamefont {Weiler}, \citenamefont {Kicanaoglu},\ and\ \citenamefont
  {Welling}}]{Cohen:2019}%
  \BibitemOpen
  \bibfield  {author} {\bibinfo {author} {\bibfnamefont {T.~S.}\ \bibnamefont
  {Cohen}}, \bibinfo {author} {\bibfnamefont {M.}~\bibnamefont {Weiler}},
  \bibinfo {author} {\bibfnamefont {B.}~\bibnamefont {Kicanaoglu}},\ and\
  \bibinfo {author} {\bibfnamefont {M.}~\bibnamefont {Welling}},\ }\bibfield
  {title} {\bibinfo {title} {{Gauge Equivariant Convolutional Networks and the
  Icosahedral {CNN}}},\ }in\ \href
  {http://proceedings.mlr.press/v97/cohen19d.html} {\emph {\bibinfo {booktitle}
  {Proceedings of the 36th International Conference on Machine Learning}}}\
  (\bibinfo {year} {2019})\ pp.\ \bibinfo {pages} {1321--1330},\ \Eprint
  {https://arxiv.org/abs/1902.04615} {arXiv:1902.04615 [cs.LG]} \BibitemShut
  {NoStop}%
\bibitem [{\citenamefont {Kanwar}\ \emph {et~al.}(2020)\citenamefont {Kanwar},
  \citenamefont {Albergo}, \citenamefont {Boyda}, \citenamefont {Cranmer},
  \citenamefont {Hackett}, \citenamefont {Racani\`ere}, \citenamefont
  {Rezende},\ and\ \citenamefont {Shanahan}}]{Kanwar:2020xzo}%
  \BibitemOpen
  \bibfield  {author} {\bibinfo {author} {\bibfnamefont {G.}~\bibnamefont
  {Kanwar}}, \bibinfo {author} {\bibfnamefont {M.~S.}\ \bibnamefont {Albergo}},
  \bibinfo {author} {\bibfnamefont {D.}~\bibnamefont {Boyda}}, \bibinfo
  {author} {\bibfnamefont {K.}~\bibnamefont {Cranmer}}, \bibinfo {author}
  {\bibfnamefont {D.~C.}\ \bibnamefont {Hackett}}, \bibinfo {author}
  {\bibfnamefont {S.}~\bibnamefont {Racani\`ere}}, \bibinfo {author}
  {\bibfnamefont {D.~J.}\ \bibnamefont {Rezende}},\ and\ \bibinfo {author}
  {\bibfnamefont {P.~E.}\ \bibnamefont {Shanahan}},\ }\bibfield  {title}
  {\bibinfo {title} {{Equivariant flow-based sampling for lattice gauge
  theory}},\ }\href {https://doi.org/10.1103/PhysRevLett.125.121601} {\bibfield
   {journal} {\bibinfo  {journal} {Phys. Rev. Lett.}\ }\textbf {\bibinfo
  {volume} {125}},\ \bibinfo {pages} {121601} (\bibinfo {year} {2020})},\
  \Eprint {https://arxiv.org/abs/2003.06413} {arXiv:2003.06413 [hep-lat]}
  \BibitemShut {NoStop}%
\bibitem [{\citenamefont {Boyda}\ \emph {et~al.}(2021)\citenamefont {Boyda},
  \citenamefont {Kanwar}, \citenamefont {Racani\`ere}, \citenamefont {Rezende},
  \citenamefont {Albergo}, \citenamefont {Cranmer}, \citenamefont {Hackett},\
  and\ \citenamefont {Shanahan}}]{Boyda:2020hsi}%
  \BibitemOpen
  \bibfield  {author} {\bibinfo {author} {\bibfnamefont {D.}~\bibnamefont
  {Boyda}}, \bibinfo {author} {\bibfnamefont {G.}~\bibnamefont {Kanwar}},
  \bibinfo {author} {\bibfnamefont {S.}~\bibnamefont {Racani\`ere}}, \bibinfo
  {author} {\bibfnamefont {D.~J.}\ \bibnamefont {Rezende}}, \bibinfo {author}
  {\bibfnamefont {M.~S.}\ \bibnamefont {Albergo}}, \bibinfo {author}
  {\bibfnamefont {K.}~\bibnamefont {Cranmer}}, \bibinfo {author} {\bibfnamefont
  {D.~C.}\ \bibnamefont {Hackett}},\ and\ \bibinfo {author} {\bibfnamefont
  {P.~E.}\ \bibnamefont {Shanahan}},\ }\bibfield  {title} {\bibinfo {title}
  {{Sampling using $SU(N)$ gauge equivariant flows}},\ }\href
  {https://doi.org/10.1103/PhysRevD.103.074504} {\bibfield  {journal} {\bibinfo
   {journal} {Phys. Rev. D}\ }\textbf {\bibinfo {volume} {103}},\ \bibinfo
  {pages} {074504} (\bibinfo {year} {2021})},\ \Eprint
  {https://arxiv.org/abs/2008.05456} {arXiv:2008.05456 [hep-lat]} \BibitemShut
  {NoStop}%
\bibitem [{\citenamefont {Abbott}\ \emph {et~al.}(2022)\citenamefont {Abbott}
  \emph {et~al.}}]{Abbott:2022zhs}%
  \BibitemOpen
  \bibfield  {author} {\bibinfo {author} {\bibfnamefont {R.}~\bibnamefont
  {Abbott}} \emph {et~al.},\ }\bibfield  {title} {\bibinfo {title}
  {{Gauge-equivariant flow models for sampling in lattice field theories with
  pseudofermions}},\ }\href {https://doi.org/10.1103/PhysRevD.106.074506}
  {\bibfield  {journal} {\bibinfo  {journal} {Phys. Rev. D}\ }\textbf {\bibinfo
  {volume} {106}},\ \bibinfo {pages} {074506} (\bibinfo {year} {2022})},\
  \Eprint {https://arxiv.org/abs/2207.08945} {arXiv:2207.08945 [hep-lat]}
  \BibitemShut {NoStop}%
\bibitem [{\citenamefont {Favoni}\ \emph {et~al.}(2022)\citenamefont {Favoni},
  \citenamefont {Ipp}, \citenamefont {M\"uller},\ and\ \citenamefont
  {Schuh}}]{Favoni:2020reg}%
  \BibitemOpen
  \bibfield  {author} {\bibinfo {author} {\bibfnamefont {M.}~\bibnamefont
  {Favoni}}, \bibinfo {author} {\bibfnamefont {A.}~\bibnamefont {Ipp}},
  \bibinfo {author} {\bibfnamefont {D.~I.}\ \bibnamefont {M\"uller}},\ and\
  \bibinfo {author} {\bibfnamefont {D.}~\bibnamefont {Schuh}},\ }\bibfield
  {title} {\bibinfo {title} {{Lattice Gauge Equivariant Convolutional Neural
  Networks}},\ }\href {https://doi.org/10.1103/PhysRevLett.128.032003}
  {\bibfield  {journal} {\bibinfo  {journal} {Phys. Rev. Lett.}\ }\textbf
  {\bibinfo {volume} {128}},\ \bibinfo {pages} {032003} (\bibinfo {year}
  {2022})},\ \Eprint {https://arxiv.org/abs/2012.12901} {arXiv:2012.12901
  [hep-lat]} \BibitemShut {NoStop}%
\bibitem [{\citenamefont {{L\"uscher}}(2004)}]{Luscher:2003qa}%
  \BibitemOpen
  \bibfield  {author} {\bibinfo {author} {\bibfnamefont {M.}~\bibnamefont
  {{L\"uscher}}},\ }\bibfield  {title} {\bibinfo {title} {{Solution of the
  Dirac equation in lattice QCD using a domain decomposition method}},\ }\href
  {https://doi.org/10.1016/S0010-4655(03)00486-7} {\bibfield  {journal}
  {\bibinfo  {journal} {Comput. Phys. Commun.}\ }\textbf {\bibinfo {volume}
  {156}},\ \bibinfo {pages} {209} (\bibinfo {year} {2004})},\ \Eprint
  {https://arxiv.org/abs/hep-lat/0310048} {arXiv:hep-lat/0310048} \BibitemShut
  {NoStop}%
\bibitem [{\citenamefont {Osaki}\ and\ \citenamefont
  {Ishikawa}(2010)}]{Osaki:2010vj}%
  \BibitemOpen
  \bibfield  {author} {\bibinfo {author} {\bibfnamefont {Y.}~\bibnamefont
  {Osaki}}\ and\ \bibinfo {author} {\bibfnamefont {K.-I.}\ \bibnamefont
  {Ishikawa}},\ }\bibfield  {title} {\bibinfo {title} {{Domain Decomposition
  method on GPU cluster}},\ }\href {https://doi.org/10.22323/1.105.0036}
  {\bibfield  {journal} {\bibinfo  {journal} {PoS}\ }\textbf {\bibinfo {volume}
  {LATTICE2010}},\ \bibinfo {pages} {036} (\bibinfo {year} {2010})},\ \Eprint
  {https://arxiv.org/abs/1011.3318} {arXiv:1011.3318 [hep-lat]} \BibitemShut
  {NoStop}%
\bibitem [{\citenamefont {Babich}\ \emph {et~al.}(2011)\citenamefont {Babich},
  \citenamefont {Clark}, \citenamefont {Joo}, \citenamefont {Shi},
  \citenamefont {Brower},\ and\ \citenamefont {Gottlieb}}]{Babich:2011np}%
  \BibitemOpen
  \bibfield  {author} {\bibinfo {author} {\bibfnamefont {R.}~\bibnamefont
  {Babich}}, \bibinfo {author} {\bibfnamefont {M.~A.}\ \bibnamefont {Clark}},
  \bibinfo {author} {\bibfnamefont {B.}~\bibnamefont {Joo}}, \bibinfo {author}
  {\bibfnamefont {G.}~\bibnamefont {Shi}}, \bibinfo {author} {\bibfnamefont
  {R.~C.}\ \bibnamefont {Brower}},\ and\ \bibinfo {author} {\bibfnamefont
  {S.}~\bibnamefont {Gottlieb}},\ }\bibfield  {title} {\bibinfo {title}
  {{Scaling Lattice QCD beyond 100 GPUs}},\ }in\ \href
  {https://doi.org/10.1145/2063384.2063478} {\emph {\bibinfo {booktitle} {{SC11
  International Conference for High Performance Computing, Networking, Storage
  and Analysis}}}}\ (\bibinfo {year} {2011})\ \Eprint
  {https://arxiv.org/abs/1109.2935} {arXiv:1109.2935 [hep-lat]} \BibitemShut
  {NoStop}%
\bibitem [{\citenamefont {Tu}\ \emph {et~al.}(2021)\citenamefont {Tu},
  \citenamefont {Clark}, \citenamefont {Jung},\ and\ \citenamefont
  {Mawhinney}}]{Tu:2021dvv}%
  \BibitemOpen
  \bibfield  {author} {\bibinfo {author} {\bibfnamefont {J.}~\bibnamefont
  {Tu}}, \bibinfo {author} {\bibfnamefont {M.~A.}\ \bibnamefont {Clark}},
  \bibinfo {author} {\bibfnamefont {C.}~\bibnamefont {Jung}},\ and\ \bibinfo
  {author} {\bibfnamefont {R.}~\bibnamefont {Mawhinney}},\ }\bibfield  {title}
  {\bibinfo {title} {{Solving DWF Dirac Equation Using Multi-splitting
  Preconditioned Conjugate Gradient with Tensor Cores on NVIDIA GPUs}},\ }in\
  \href {https://doi.org/10.1145/3468267.3470613} {\emph {\bibinfo {booktitle}
  {{PASC '21: Proceedings of the Platform for Advanced Scientific Computing
  Conference}}}}\ (\bibinfo {year} {2021})\ pp.\ \bibinfo {pages} {1--11},\
  \Eprint {https://arxiv.org/abs/2104.05615} {arXiv:2104.05615 [hep-lat]}
  \BibitemShut {NoStop}%
\bibitem [{\citenamefont {Wilson}(1974)}]{Wilson:1974sk}%
  \BibitemOpen
  \bibfield  {author} {\bibinfo {author} {\bibfnamefont {K.~G.}\ \bibnamefont
  {Wilson}},\ }\bibfield  {title} {\bibinfo {title} {{Confinement of Quarks}},\
  }\href {https://doi.org/10.1103/PhysRevD.10.2445} {\bibfield  {journal}
  {\bibinfo  {journal} {Phys. Rev. D}\ }\textbf {\bibinfo {volume} {10}},\
  \bibinfo {pages} {2445} (\bibinfo {year} {1974})}\BibitemShut {NoStop}%
\bibitem [{\citenamefont {Sheikholeslami}\ and\ \citenamefont
  {Wohlert}(1985)}]{Sheikholeslami:1985ij}%
  \BibitemOpen
  \bibfield  {author} {\bibinfo {author} {\bibfnamefont {B.}~\bibnamefont
  {Sheikholeslami}}\ and\ \bibinfo {author} {\bibfnamefont {R.}~\bibnamefont
  {Wohlert}},\ }\bibfield  {title} {\bibinfo {title} {{Improved continuum limit
  lattice action for QCD with Wilson fermions}},\ }\href
  {https://doi.org/10.1016/0550-3213(85)90002-1} {\bibfield  {journal}
  {\bibinfo  {journal} {Nucl. Phys. B}\ }\textbf {\bibinfo {volume} {259}},\
  \bibinfo {pages} {572} (\bibinfo {year} {1985})}\BibitemShut {NoStop}%
\bibitem [{\citenamefont {Saad}(1993)}]{FGMRES}%
  \BibitemOpen
  \bibfield  {author} {\bibinfo {author} {\bibfnamefont {Y.}~\bibnamefont
  {Saad}},\ }\bibfield  {title} {\bibinfo {title} {{A Flexible Inner-Outer
  Preconditioned GMRES Algorithm}},\ }\href {https://doi.org/10.1137/0914028}
  {\bibfield  {journal} {\bibinfo  {journal} {SIAM J. Sci. Comput.}\ }\textbf
  {\bibinfo {volume} {14}},\ \bibinfo {pages} {461} (\bibinfo {year}
  {1993})}\BibitemShut {NoStop}%
\bibitem [{\citenamefont {Shamir}(1993)}]{Shamir:1993zy}%
  \BibitemOpen
  \bibfield  {author} {\bibinfo {author} {\bibfnamefont {Y.}~\bibnamefont
  {Shamir}},\ }\bibfield  {title} {\bibinfo {title} {{Chiral fermions from
  lattice boundaries}},\ }\href {https://doi.org/10.1016/0550-3213(93)90162-I}
  {\bibfield  {journal} {\bibinfo  {journal} {Nucl. Phys. B}\ }\textbf
  {\bibinfo {volume} {406}},\ \bibinfo {pages} {90} (\bibinfo {year} {1993})},\
  \Eprint {https://arxiv.org/abs/hep-lat/9303005} {arXiv:hep-lat/9303005}
  \BibitemShut {NoStop}%
\bibitem [{\citenamefont {Furman}\ and\ \citenamefont
  {Shamir}(1995)}]{Furman:1994ky}%
  \BibitemOpen
  \bibfield  {author} {\bibinfo {author} {\bibfnamefont {V.}~\bibnamefont
  {Furman}}\ and\ \bibinfo {author} {\bibfnamefont {Y.}~\bibnamefont
  {Shamir}},\ }\bibfield  {title} {\bibinfo {title} {{Axial symmetries in
  lattice QCD with Kaplan fermions}},\ }\href
  {https://doi.org/10.1016/0550-3213(95)00031-M} {\bibfield  {journal}
  {\bibinfo  {journal} {Nucl. Phys. B}\ }\textbf {\bibinfo {volume} {439}},\
  \bibinfo {pages} {54} (\bibinfo {year} {1995})},\ \Eprint
  {https://arxiv.org/abs/hep-lat/9405004} {arXiv:hep-lat/9405004} \BibitemShut
  {NoStop}%
\bibitem [{\citenamefont {Kingma}\ and\ \citenamefont {Ba}(2014)}]{ADAM}%
  \BibitemOpen
  \bibfield  {author} {\bibinfo {author} {\bibfnamefont {D.~P.}\ \bibnamefont
  {Kingma}}\ and\ \bibinfo {author} {\bibfnamefont {J.}~\bibnamefont {Ba}},\
  }\href@noop {} {\bibinfo {title} {Adam: A method for stochastic
  optimization}} (\bibinfo {year} {2014}),\ \Eprint
  {https://arxiv.org/abs/1412.6980} {arXiv:1412.6980 [cs.LG]} \BibitemShut
  {NoStop}%
\bibitem [{\citenamefont {{L\"uscher}}(2007)}]{Luscher:2007se}%
  \BibitemOpen
  \bibfield  {author} {\bibinfo {author} {\bibfnamefont {M.}~\bibnamefont
  {{L\"uscher}}},\ }\bibfield  {title} {\bibinfo {title} {{Local coherence and
  deflation of the low quark modes in lattice QCD}},\ }\href
  {https://doi.org/10.1088/1126-6708/2007/07/081} {\bibfield  {journal}
  {\bibinfo  {journal} {JHEP}\ }\textbf {\bibinfo {volume} {07}},\ \bibinfo
  {pages} {081}},\ \Eprint {https://arxiv.org/abs/0706.2298} {arXiv:0706.2298
  [hep-lat]} \BibitemShut {NoStop}%
\end{thebibliography}%

\clearpage
\end{document}